\documentclass[aip,pop,reprint]{revtex4-1}

\usepackage{graphicx}                       
\usepackage[outdir=./]{epstopdf}
\usepackage{amsmath}
\usepackage{amssymb}
\usepackage{amsthm}
\usepackage{dsfont}       
\usepackage[usenames,dvipsnames,table]{xcolor}
\usepackage[format=plain, font=small, labelfont=bf, justification = raggedright]{caption}
\usepackage{array}  
\usepackage{overpic}
\usepackage{placeins} 
\usepackage{fancyhdr}
\usepackage{empheq} 
\usepackage{colortbl}
\usepackage{multirow} 
\usepackage{hyperref} 
\hypersetup{colorlinks = true,linkcolor = blue, breaklinks = true}

\newcommand{\eq}[1]{(\ref{#1})}
\newcommand{\Eq}[1]{Eq.~\eq{#1}}
\newcommand{\Eqs}[1]{Eqs.~\eq{#1}}
\newcommand{\Fig}[1]{Fig.~\ref{#1}}
\newcommand{\Sec}[1]{Sec.~\ref{#1}}

\renewcommand{\Ref}[1]{Ref.~\onlinecite{#1}}

\newcommand{\App}[1]{Appendix~\ref{#1}}

\newcommand{\eg}{{e.g., }}
\newcommand{\ie}{{i.e., }}

\newcommand{\mc}[1]{\mathcal{#1}}

\newcommand{\Vect}[1]{{\boldsymbol{\rm #1}}}

\newcommand{\fourier}[1]{\smash{\widetilde{#1}}}

\newcommand{\pd}[1]{\partial_{#1}}

\newcommand{\dd}{\mathrm{d}}

\DeclareMathOperator{\airyA}{Ai}

\DeclareMathOperator{\sinc}{sinc}
\DeclareMathOperator{\erf}{erf}

\newcommand{\Symb}[1]{\mc{#1}}

\renewcommand{\Re}{\textrm{Re}}

\newcommand{\airySKIN}{\delta_a}
\newcommand{\airyLEN}{L}
\newcommand{\compLEN}{\airyLEN_c}
\newcommand{\airyNU}{\nu}
\newcommand{\airyARG}{C}
\newcommand{\focal}{f}
\newcommand{\waist}{\varpi}
\newcommand{\compFOCAL}{\focal_c}
\newcommand{\normTWO}{\mc{N}}
\newcommand{\normONE}{\overline{\mc{N}}}
\newcommand{\hUMB}{\text{U}_\text{H}}
\newcommand{\hUMBKER}{\mc{U}_\text{H}}
\newcommand{\width}{W}
\newcommand{\aperFUNC}{A}

\begin{document}
\setlength{\parskip}{0pt}
\setlength{\belowcaptionskip}{0pt}


\title{On the intensity of focused waves near turning points}
\author{N. A. Lopez}
\affiliation{Department of Astrophysical Sciences, Princeton University, Princeton, New Jersey 08544, USA}
\affiliation{Rudolf Peierls Centre for Theoretical Physics, University of Oxford,
Oxford OX1 3PU, UK}
\author{E. Kur}
\affiliation{Lawrence Livermore National Laboratory, Livermore, California 94551, USA}
\author{D. J. Strozzi}
\affiliation{Lawrence Livermore National Laboratory, Livermore, California 94551, USA}

\begin{abstract}
A wave near an isolated turning point is typically assumed to have an Airy function profile with respect to the separation distance. This description is incomplete, however, and is insufficient to describe the behavior of more realistic wavefields that are not simple plane waves. Asymptotic matching to a prescribed incoming wavefield generically introduces a phasefront curvature term that changes the characteristic wave behavior from the Airy function to that of the hyperbolic umbilic function. This function, which is one of the seven classic `elementary' functions from catastrophe theory along with the Airy function, can be understood intuitively as the solution for a linearly focused Gaussian beam propagating in a linearly varying density profile, as we show. The morphology of the caustic lines that govern the intensity maxima of the diffraction pattern as one alters the density lengthscale of the plasma, the focal length of the incident beam, and also the injection angle of the incident beam are presented in detail. This morphology includes a Goos--H\"anchen shift and focal shift at oblique incidence that do not appear in a reduced ray-based description of the caustic. The enhancement of the intensity swelling factor for a focused wave compared to the typical Airy solution is highlighted, and the impact of finite lens aperture is discussed. Collisional damping and finite beam waist are included in the model and appear as complex components to the arguments of the hyperbolic umbilic function. The observations presented here on the behavior of waves near turning points should aid the development of improved reduced wave models to be used, for example, in designing modern nuclear fusion experiments.
\end{abstract}

\maketitle

\pagestyle{fancy}
\lhead{Lopez, Kur, \& Strozzi}
\rhead{Focused waves near turning points}
\thispagestyle{empty}

\section{Introduction}

Basic wave physics is central to the development of controlled thermonuclear fusion. Indeed, a key component to the fusion milestones recently obtained at the National Ignition Facility (NIF)~\cite{AbuShawareb22,Zylstra22} was leveraging the basic nonlinear optical process of cross-beam energy transfer (CBET) to maintain drive symmetry~\cite{Michel09,Michel10}. That said, there still remain open questions regarding waves in fusion-relevant plasmas. One such question is the amount of reflection losses (\ie glint) of an incident laser beam off the ablating hohlraum wall, an issue that was the focus of a recent experimetal campaign on NIF~\cite{Lemos22} for its possible connection to explaining the drive-deficit problem~\cite{Jones17}. Being able to predict glint is paramount to future experimental performance because at sufficiently high intensities, the glint light may get nonlinearly amplified via CBET with lasers incident from the opposite entrance hole~\cite{Turnbull15} or even with light from the original laser~\cite{Colaitis19b,Kur20DPP} to increase the amount lost. Moreover, researchers have recently begun to wonder~\cite{Lopez21DPP} whether the reflection physics might be modified by the speckle hot spots that are necessarily introduced to the NIF laser by the smoothing phase plates~\cite{Spaeth17}, since these are not currently accounted for in many inline laser modules. Hence, understanding the intensity profile of general wavefields near turning points has a renewed importance in fusion research. 

For a plane wave incident on an isolated turning point (or turning plane in multiple dimensions), the intensity profile is well-known to be given by Airy's function~\cite{Olver10a}. For nonplanar wavefields, one can sometimes perform an asymptotic matching onto the Airy function and its derivative~\cite{Olver10a,Kravtsov93}, but this can often obscure some of the key properties of the true solution. Indeed, the Airy function is merely the simplest member of a large hierarchy of functions (the so-called diffraction integrals of catastrophe theory~\cite{Kravtsov93,Poston96,Berry80b}) that can be used to describe wave behavior near critical points; since all members of this hierarchy contain the Airy function as a limiting behavior, it stands to reason that more general behavior might be more accurately and compactly captured by using higher order catastrophe functions. 

Here it is shown that the behavior of a general wavefield near an isolated turning point can be compactly expressed as an integral mapping whose kernel is the hyperbolic umbilic function. The integral mapping takes the form of a convolution at normal incidence. As anticipated, the hyperbolic umbilic function is a higher member of the catastrophe hierarchy that allows non-plane-wave behavior near turning points (specifically, phasefront curvature) to be concisely described. By itself, the hyperbolic umbilic function also describes the solution for a Gaussian focused beam incident on a turning point. Since this function is not common in plasma physics, considerable space is dedicated to describing the morphology of the hyperbolic umbilic function as parameters of the problem are altered. The effects of dissipation and finite aperture width are also discussed. Due to their general nature, the results presented here will have applications beyond laser fusion experiments; for example, they may be useful for magnetic fusion researchers attempting to heat overdense plasmas via mode-conversion methods~\cite{Igami06,Urban11,Taylor15,Lopez18b}, or attempting to measure turbulent fluctuations via Doppler backscattering~\cite{RuizRuiz22,HallChen22}.

This paper is organized as follows. In \Sec{sec:setup} the basic problem is set up. In \Sec{sec:solution} the general solution for an arbitrary incident wavefield is obtained, which can be considered the main result of this work. In \Sec{sec:cases} the special cases of a plane wave and a focused Gaussian wave (with and without aperture) are studied in detail as a means of understanding the general result presented in the previous section. Lastly, in \Sec{sec:concl} the main results are summarized. Additional discussions are provided in appendices.


\section{Problem setup}
\label{sec:setup}

\begin{figure}
    \includegraphics[width=0.75\linewidth]{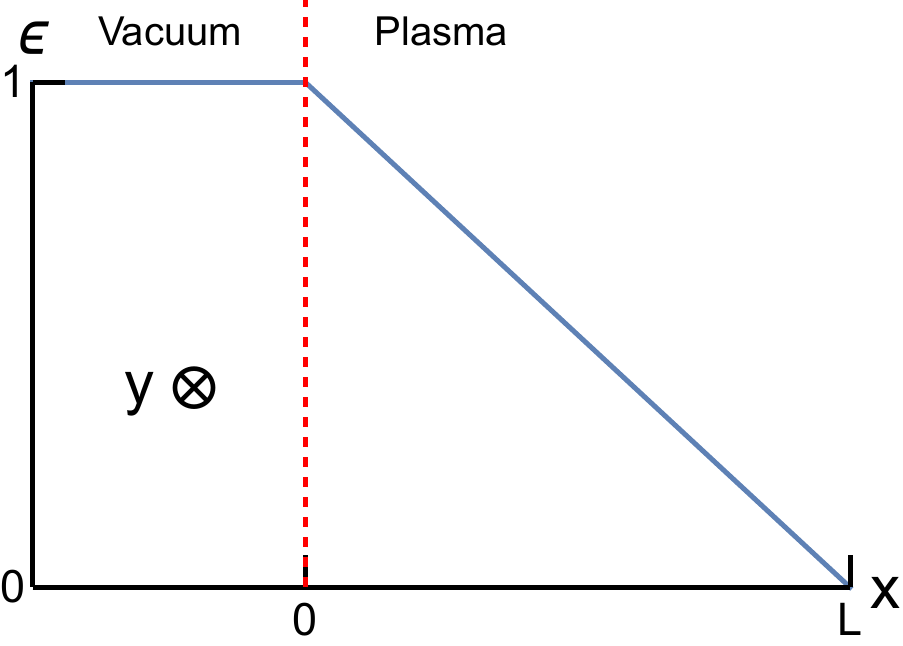}
    \caption{The plasma dielectric function is assumed to be linear in $x$, with vacuum-plasma boundary at $x = 0$, and cutoff at $x = \airyLEN$, and homogeneous in $y$, which, as noted, points into the page of the figure.}
    \label{fig:epsilon}
\end{figure}

Let us consider a beam propagating in two dimensions ($2$-D)%
\footnote{The following analysis can be readily generalized to arbitrary number of dimensions through only cosmetic modifications} 
in a plasma that varies only in one direction, which we take to be $x$, with $y$ being the remaining spatial direction. Since we are interested in the wavefield behavior near a turning point, we adopt a linear approximation of the plasma dielectric function (see \Fig{fig:epsilon}):
\begin{equation}
	\epsilon(x \ge 0) = 
	1 - \frac{x}{\airyLEN ( 1 + i \airyNU )}
	,
\end{equation}

\noindent where $\airyLEN$ is a constant length scale and $\airyNU \ge 0$ is a constant dimensionless damping coefficient. (Note that all numerical plots will have $\nu = 0$.) Hence, $x = 0$ corresponds to the vacuum-plasma interface (although it can be made to correspond to a more general boundary condition by setting $\epsilon(0) = \epsilon_0 > 0$ rather than unity). Lastly, let us assume the beam oscillates monochromatically in time and the plasma profile is stationary. Hence, we can partition the total wavefield $\Vect{E}$ as
\begin{equation}
	\Vect{E}(x, y, t) = \psi(x, y) \exp\left(- 2 \pi i \frac{ ct }{\lambda} \right) \hat{\Vect{e}} + c.c.
	,
\end{equation}

\noindent where $c$ is the speed of light in vacuum, $\lambda$ is the vacuum wavelength of the launched beam, and $\hat{\Vect{e}}$ is the polarization vector. We shall further take $\Vect{E}$ to be $s$-polarized such that $\hat{\Vect{e}}$ plays no role in the propagation dynamics and can be discarded.


\section{General solution}
\label{sec:solution}

\subsection{Local solution near the turning point}

Near the turning point for $x \ge 0$, the wavefield satisfies the Helmholtz equation
\begin{equation}
	\left(
		\pd{x}^2 
		+ \pd{y}^2
		+
		\frac{\compLEN - x}{\airySKIN^3}
	\right)\psi(x, y) = 0
	,
	\label{eq:airyEQ}
\end{equation}

\noindent where we have introduced the complex lengthscale
\begin{equation}
    \compLEN \doteq \airyLEN(1 + i \airyNU)
    ,
\end{equation}

\noindent along with the complex Airy skin depth
\begin{equation}
	\airySKIN \doteq
	\sqrt[3]{\frac{\compLEN \lambda^2}{4 \pi^2}}
	.
\end{equation}

\noindent The conditions $\airyLEN > 0$ and $\airyNU \ge 0$ restrict $\compLEN$ to the first quadrant of the complex plane, \ie $\arg(\compLEN) \in [0, \pi/2)$.

To proceed, let us apply a shifted Fourier transform (FT) in $y$. Our FT convention is as follows:%
\footnote{All integrals range from $-\infty$ to $+\infty$ unless explicitly stated.}%
\begin{subequations}
    \label{eq:shiftFT}
	\begin{align}
		\fourier{f}(k_y) 
		&= \int \frac{\dd y}{\sqrt{2\pi}} 
		f(y) \exp\left[ 
			- i y \left(k_y + \frac{2\pi}{\lambda} \sin \theta \right)
		\right]
		, \\
		f(y) 
		 &= \int \frac{\dd k_y}{\sqrt{2\pi}} 
		 \fourier{f}(k_y) \exp\left[ 
			i y \left(k_y + \frac{2\pi}{\lambda} \sin \theta \right)
		\right]
		 ,
	\end{align}
\end{subequations}

\noindent where we have transformed out the mean wavevector $2\pi \sin \theta / \lambda$, which is assumed to be the predominant direction of $k_y$ for oblique propagation. Applying \Eq{eq:shiftFT} to \Eq{eq:airyEQ} then yields
\begin{equation}
	\left[
		\airySKIN^2 \pd{x}^2
		- \airyARG\left(x, k_y + \frac{2\pi}{\lambda} \sin \theta \right)
	\right]
	\fourier{\psi}(x, k_y)
	= 0
	,
	\label{eq:transformEQ}
\end{equation}

\noindent where we have introduced the cutoff function
\begin{equation}
	\airyARG(x, k_y)
	\doteq
	\airySKIN^2 \,k_y^2
	+ \frac{x - \compLEN}{\airySKIN}
	.
    \label{eq:cutoffFUNC}
\end{equation}

\noindent Note that $\text{Re}(C) = 0$ defines the turning point for a plane wave propagating obliquely with transverse wavevector $k_y$. The solution to \Eq{eq:transformEQ} which remains regular as $x \to + \infty$ is given by%
\footnote{The restriction $\arg(\compLEN) \in [0, \pi/2)$ in turn restricts $\arg(\airyARG) \in [0, -\pi/6)$ as $x \to +\infty$; hence, no Stokes' phenomenon correction to \Eq{eq:airyFT} are needed as these would require $|\arg(C)| \ge \pi/3$~\cite{Heading62a}.}%
\begin{equation}
	\fourier{\psi}(x, k_y)
	=
	\frac{
		\airyA\left[
			\airyARG\left(x, k_y + \frac{2\pi}{\lambda} \sin \theta \right)
		\right]
	}{
		\airyA\left[
			\airyARG\left(0, k_y + \frac{2\pi}{\lambda} \sin \theta \right)
		\right]
	}
	\,
	\fourier{\psi}(0, k_y)
	,
	\label{eq:airyFT}
\end{equation}

\noindent where $\airyA$ is the Airy function~\cite{Olver10a}. The general solution for $x \ge 0$ can then be obtained by performing an inverse FT.


\subsection{Asymptotic matching at plasma-vacuum boundary}

Computing \Eq{eq:airyFT} requires knowing the total wavefield $\psi(0, y)$, which includes the interference between incoming and reflected components. This is difficult to construct when the reflected wavefield is itself the object of inquiry. An analogous equation to \Eq{eq:airyFT} that depends only on the incoming field can be obtained via asymptotic matching as follows.

Suppose that the inverse-FT integral to obtain $\psi(x,y)$ from \Eq{eq:airyFT} is negligible beyond some characteristic maximum wavevector $k_\text{max}$%
\footnote{This might be either because for $|k_y| > k_\text{max}$ the spectrum of $\psi$ decays rapidly to zero or the integral itself becomes increasingly oscillatory and averages to zero.}. %
If the input plane $x = 0$ is located asymptotically far from the turning point for the maximum wavevector, i.e.,
\begin{equation}
	- \Re\left[ \airyARG\left(0, k_\text{max} + \frac{2\pi}{\lambda} \sin \theta \right) \right] \gg 1
	,
	\label{eq:asympAIRYineq}
\end{equation}

\noindent then $\airyARG$ at $x = 0$ is large and negative for all $|k_y| \le k_\text{max}$, and we can use the asymptotic approximation%
\footnote{Again, no Stokes' phenomenon corrections are needed because $\arg(-C) \approx \arg(\compLEN/\airySKIN) \in [0, \pi/3)$ is within the wedge of validity $\arg(\zeta) \in (-2\pi/3, 2\pi/3)$ for the standard asymptotic expansion of $\airyA(-\zeta)$~\cite{Olver10a}.}%
\begin{equation}
	\airyA\left[ \airyARG\left(0 \right) \right]
	\approx
	\frac{
		\cos\left\{
			\frac{2}{3}
			\left[ -\airyARG\left(0 \right) \right]^{3/2}
			- \frac{\pi}{4}
		\right\}
	}{
		\sqrt{\pi}
		\sqrt[4]{- \airyARG\left(0 \right) }
	} 
	,
	\label{eq:asympAIRY}
\end{equation}

\noindent where here and in the following we have suppressed the second argument to $\airyARG$ for brevity; it is understood to be $k_y + \frac{2\pi}{\lambda} \sin \theta$. Using \Eq{eq:asympAIRY} in \Eq{eq:airyFT} allows the incoming component to be isolated as
\begin{equation}
	\fourier{\psi}_\text{in}(0, k_y)
    =
	\frac{
		\exp\left\{
			- i \frac{2}{3}
			\left[ - \airyARG(0) \right]^{3/2}
		\right\}
		\fourier{\psi}(x, k_y )
	}{
		\sqrt{- 4\pi i}
		\sqrt[4]{- \airyARG(0)}
		\airyA\left[ \airyARG(x) \right]
	}
	.
	\label{eq:asymIN}
\end{equation}

Note that the asymptotic matching condition \eq{eq:asympAIRYineq} can also be understood as a paraxial requirement on the \emph{incoming} field (but not necessarily the entire field), as anticipated by the shifted FT used in \Eq{eq:shiftFT}: 
\begin{equation}
	0 \le k_\text{max} \ll \frac{2\pi}{\lambda} \left( \sqrt{1 - \frac{\airySKIN}{\airyLEN}} - \sin \theta \right)
	.
	\label{eq:smallKY}
\end{equation}

\noindent The necessary condition
\begin{equation}
	\sin \theta \le \sqrt{1 - \frac{\airySKIN}{\airyLEN}}
	\label{eq:necessary}
\end{equation}

\noindent then places a limit on the maximum angle of incidence describable by our model. Physically, these two conditions \eq{eq:smallKY} and \eq{eq:necessary} arise because oblique propagation with transverse wavevector $k_y$ shifts the turning point closer to the input plane by $\airySKIN^3 k_y^2$, as seen from \Eq{eq:cutoffFUNC}. Equations \eq{eq:necessary} and \eq{eq:smallKY} therefore state that the turning points for the mean wavevector $2\pi \sin\theta/\lambda$ and for all deviations from the mean wavevector contained within the incoming wave spectrum are also located far from the input plane. Note also that for shallow incidence ($\theta$ large) when our matching scheme fails, one can instead perform the asymptotic matching spatially rather than spectrally because at such angles the overlap region between the incoming and reflected wavefields is small. At sufficiently shallow incidence, one might even be able to propagate $\psi$ according to the paraxial wave equation, with the general solution for a linear density gradient provided in \Ref{Lopez22t}. We shall not pursue such generalizations here.

Further simplifications to \Eq{eq:asymIN} consistent with the paraxial approximation \eq{eq:smallKY} can be performed%
\footnote{Formally, \Eqs{eq:approx} require $k_y \ll \frac{2\pi}{\lambda} \cos \theta \times \textrm{min}\left(1, \frac{\cot \theta}{2}\right)$.}. %
First, we make a slow-envelope approximation such that
\begin{subequations}
	\label{eq:approx}%
    \begin{equation}
	    \label{eq:approx1}
		\sqrt[4]{-\airyARG\left(0\right) }
		\approx \sqrt[6]{ \frac{2 \pi \compLEN}{\lambda} \cos^3 \theta }
		.
	\end{equation}

	\noindent However, we shall retain the $k_y$ dependence in the phase:
	\begin{align}
		\label{eq:approx2}
        \left[ - \airyARG\left(0\right) \right]^{3/2}
		&\approx
		\frac{2 \pi \compLEN}{\lambda} \cos^3 \theta
		- \frac{3}{2} \compLEN k_y \sin 2 \theta
		\nonumber\\
		&\hspace{23mm}
		- \frac{3 \lambda \compLEN}{4\pi} k_y^2 \frac{\cos 2 \theta}{\cos \theta}
		.
	\end{align}
\end{subequations}

\noindent Note that the approximations given by \Eqs{eq:asympAIRY} and \eq{eq:approx} only alter the initial conditions of the Fourier-space solution \eq{eq:airyFT} to the Helmholtz equation and hence preserve the `exactness' of the solution. Said differently, the $\fourier{\psi}$ obtained via \Eq{eq:asymIN} exactly solves \Eq{eq:transformEQ} regardless the functional form of $\airyARG(0)$. These approximations instead alter which exact solution a given $\psi_\text{in}$ is mapped to. 

Performing an inverse FT to \Eq{eq:asymIN} and using \Eqs{eq:approx} therefore yields the matched solution%
\begin{widetext}
    \begin{subequations}
        \label{eq:AIRYsol}%
        \begin{align}
           	\psi(x, y)
	        &\approx 
            \normTWO
          	\int \dd y' \, 
	        \psi_\text{in}( 0,y')
          	\hUMB
	        \left(
		        \sqrt[3]{3} \frac{x - \compLEN}{\airySKIN}
          		,
	        	\frac{ 
		        	(y - y') \cos \theta
			        - 2 \compLEN \sin^3 \theta
          		}{\sqrt[6]{3} \, \airySKIN \cos \theta}
	        	,
    		    - \frac{\lambda \compLEN}{2\pi \sqrt[3]{3} \, \airySKIN^2 } \frac{\cos 2 \theta}{\cos \theta}
           	\right)
    	    ,
       \end{align}
    
        \noindent or equivalently in terms of $\fourier{\psi}_\text{in}$ instead of $\psi_\text{in}$,
       \begin{align}
            \psi(x, y)
            &\approx
          	\normTWO \sqrt{2 \pi}
        	\int \dd u \, \dd v
	        \,
    	    \fourier{\psi}_\text{in}\left( 0, \frac{u}{\sqrt[6]{3} \, \airySKIN} - \frac{2\pi}{\lambda} \sin \theta \right)
    	    \hUMBKER
		    \left(
			    u, v,
			    \sqrt[3]{3} \frac{x - \compLEN}{\airySKIN},
			    \frac{ 
		    	    y \cos \theta
			        - 2 \compLEN \sin^3 \theta
    		    }{\sqrt[6]{3} \, \airySKIN \cos \theta},
			    - \frac{\lambda \compLEN}{2\pi \sqrt[3]{3} \, \airySKIN^2 } \frac{\cos 2 \theta}{\cos \theta}
		    \right)
		    ,
        \end{align}
    \end{subequations}

    \noindent where the single integral over $y'$ has been replaced by a double integral over $u$ and $v$ (which may in fact be easier to solve at times), the normalization constant is given as
    \begin{align}
        \normTWO \doteq
        \frac{\sqrt[6]{3}}{2\pi \sqrt{i \pi} \airySKIN} 
        \sqrt[6]{ \frac{2 \pi \compLEN}{\lambda} \cos^3 \theta }
        \exp\left(i \frac{4 \pi \compLEN}{3 \lambda} \frac{2 - \sin^2 \theta \cos2 \theta }{2\cos \theta} \right)
        ,
        \label{eq:originalNORM}
    \end{align}
\end{widetext}

\noindent and we have introduced the function $\hUMB$, defined as
\begin{subequations}
	\begin{align}
		\hUMB(t_1, t_2, t_3)
		&\doteq
		\int \dd u \, \dd v \, \,
		\hUMBKER
		\left(
			u, v,
			t_1, t_2, t_3
		\right)
		,
        \label{eq:hyperUMB}
	\end{align}
	
	\noindent and the function $\hUMBKER$, defined as
	\begin{align}
		\hUMBKER
		\left(
			u, v,
			t_1, t_2, t_3
		\right)
		&\doteq
		\exp(
			i u^2 \, v
			+ i v^3
			+ i t_3 \, u^2
			\nonumber\\
			&\hspace{19mm}
			+ i t_2 u
			+ i t_1 v 
		)
		,
		\label{eq:quasiHU}
	\end{align}
\end{subequations}

\noindent as the standard $D_4^+$ hyperbolic umbilic catastrophe function~\cite{Poston96} and the hyperbolic umbilic density function, respectively. $\hUMB$ is one of the famous seven elementary diffraction catastrophes~\cite{Thom75,Berry76} (the simplest of which being the Airy function). Note also that $[\hUMB(t_1, t_2, t_3)]^* = \hUMB(t_1, t_2, -t_3)$ and $\hUMB(t_1, -t_2, t_3) = \hUMB(t_1, t_2, t_3)$ when all parameters are real. We shall discuss $\hUMB$ in more detail in the following section and in \App{app:Humbilic}.

Before doing so, however, it is worthwhile to emphasize the advantages of introducing $\hUMB$ into the analysis. The main advantage is the structural stability of $\hUMB$. This feature both justifies the dropping of higher-order terms in \Eq{eq:approx2} by invoking strong $4$-determinancy~\cite{Poston96} and suggests that the general phenomenon described by \Eq{eq:AIRYsol} will persist even if the problem setup changes moderately, \ie replacing the linear plasma profile by an exponential profile more typical of a freely expanding plasma. As is apparent from \Eq{eq:AIRYsol}, the structural stability of $\hUMB$ also manifests as an `invariance' of sorts with respect to injection angle: the injection angle appears simply as a parameter within the arguments of $\hUMB$ that will cause the field profile to be translated, sheared, etc., but will not change the general functional behavior of the solution. [In fact, the angle-dependent terms in \Eq{eq:AIRYsol} can be identified as gradient-index analogues of the Goos--H\"anchen and focal shifts~\cite{McGuirk77}.] This is analogous to the `invariance' of the Airy function (which is also structurally stable) to the injection angle as implied by \Eq{eq:airyFT}. However, as we shall now show, the $\hUMB$ representation is superior over the standard Airy representation due to its ability to compactly describe the fields that result from incident \textit{beams} instead of plane waves.


\section{Special cases}
\label{sec:cases}

\subsection{Special case: plane wave}
\label{sec:plane}

As a sanity check, let us first confirm that \Eq{eq:AIRYsol} recovers the correct solution when $\psi_\text{in}$ corresponds to a plane wave. Setting 
\begin{equation}
    \psi_\text{in}(0,y) = E_0 \exp\left(i \frac{2\pi y}{\lambda} \sin \theta \right)
\end{equation}

\noindent (with $E_0$ a constant) in \Eq{eq:AIRYsol} yields
\begin{align}
	\psi(x, y)
	&=
	\frac{2 \pi E_0}{ \sqrt{i \pi}}
	\sqrt[6]{ \frac{2 \pi \compLEN}{\lambda} \cos^3 \theta } \,
	\airyA\left[
		\airyARG\left(x, \frac{2\pi}{\lambda} \sin \theta \right)
	\right]
	\nonumber\\
	&\hspace{4mm}\times
	\exp\left(
		i \frac{2 \pi y}{\lambda} \sin \theta
		+ i \frac{4 \pi \compLEN}{3 \lambda}
		\cos^3 \theta
	\right)
	.
	\label{eq:obliqueAIRY}
\end{align}

\noindent as desired. (Note that the constants ensure the incoming component of $\airyA$ has amplitude $E_0$.) Since $\max(\airyA) \approx 0.5$ along the real line, one can estimate the swelling factor for the intensity of \Eq{eq:obliqueAIRY} when $\airyNU = 0$ as
\begin{equation}
	\left| 
		\frac{\psi_\text{max}}{\psi_\text{in}} 
	\right|^2 \sim 
	\left(
		\frac{2 \pi \airyLEN}{\lambda}
	\right)^{1/3}
	\pi \cos \theta
	,
	\label{eq:swellAIRY}
\end{equation}

\noindent which is in agreement with known results~\cite{Myatt17}. Importantly, the power-law scaling of the swelling factor \eq{eq:swellAIRY} with respect to $1/\lambda$ ($1/3$) is equal to twice the singularity index of the fold catastrophe function~\cite{Berry80b,Olver10a} ($1/6$), as expected.


\subsection{Special case: focused Gaussian beam}
\label{sec:focused}

To develop more intuition for what $\hUMB$ is in \Eq{eq:AIRYsol}, let us also consider the case when $\psi_\text{in}$ corresponds to a focused wave with a Gaussian envelope:
\begin{equation}
	\psi_\text{in}(0,y) = 
	\frac{E_0 \cos \theta}{\sqrt{\compFOCAL/\focal}}
	\exp\left( 
		i \frac{2 \pi y}{\lambda} \sin \theta
		- i \frac{\pi y^2}{\lambda \compFOCAL}  \cos^2 \theta
	\right)
	,
	\label{eq:psiINhumb}
\end{equation}

\noindent which is the field behavior of a weakly focused Gaussian beam within a Rayleigh range of the focal plane~\cite{Siegman86}. Here $\compFOCAL$ is the complex beam parameter~\cite{Siegman86} (equivalently, a complex focal length) whose real and imaginary parts are $\compFOCAL \doteq \focal + i \waist$, with $\focal$ being the focal length and $\waist \ge 0$ parameterizing the beam waist (with $\waist = 0$ being a focused plane wave). Equation \eq{eq:AIRYsol} then yields
\begin{widetext}
\begin{align}
	\psi(x, y)
	&=
	\normONE \,
	\hUMB
	\left(
		\sqrt[3]{3} \frac{x - \compLEN}{\airySKIN},
		\frac{ 
			y \cos \theta
			- 2 \compLEN \sin^3 \theta
			- \compFOCAL \tan \theta
		}{\sqrt[6]{3} \, \airySKIN \cos \theta},
		\lambda \frac{
			\compFOCAL 
			- 2 \compLEN \cos 2 \theta \cos \theta
		}{4 \pi \sqrt[3]{3} \, \airySKIN^2 \cos^2 \theta}
	\right)
	,
	\label{eq:gaussSOL}
\end{align}

\noindent where we have introduced the normalization constant
\begin{align}
	\normONE
	\doteq
	E_0 
	\frac{ \sqrt[6]{3} \sqrt{\lambda \focal }  }{2\pi i \sqrt{\pi} \airySKIN}
	\sqrt[6]{ \frac{2 \pi \compLEN}{\lambda} \cos^3 \theta } \,
	\exp\left(
		i \frac{4 \pi \compLEN}{3 \lambda} 
		\frac{2 - \sin^2 \theta \cos2 \theta }{2\cos \theta}
		+ i \frac{\pi \compFOCAL}{\lambda} \tan^2 \theta
	\right)
	.
	\label{eq:normONEdef}
\end{align}
\end{widetext}

\begin{figure}
	\includegraphics[width=0.8\linewidth,trim={4mm 14mm 5mm 21mm},clip]{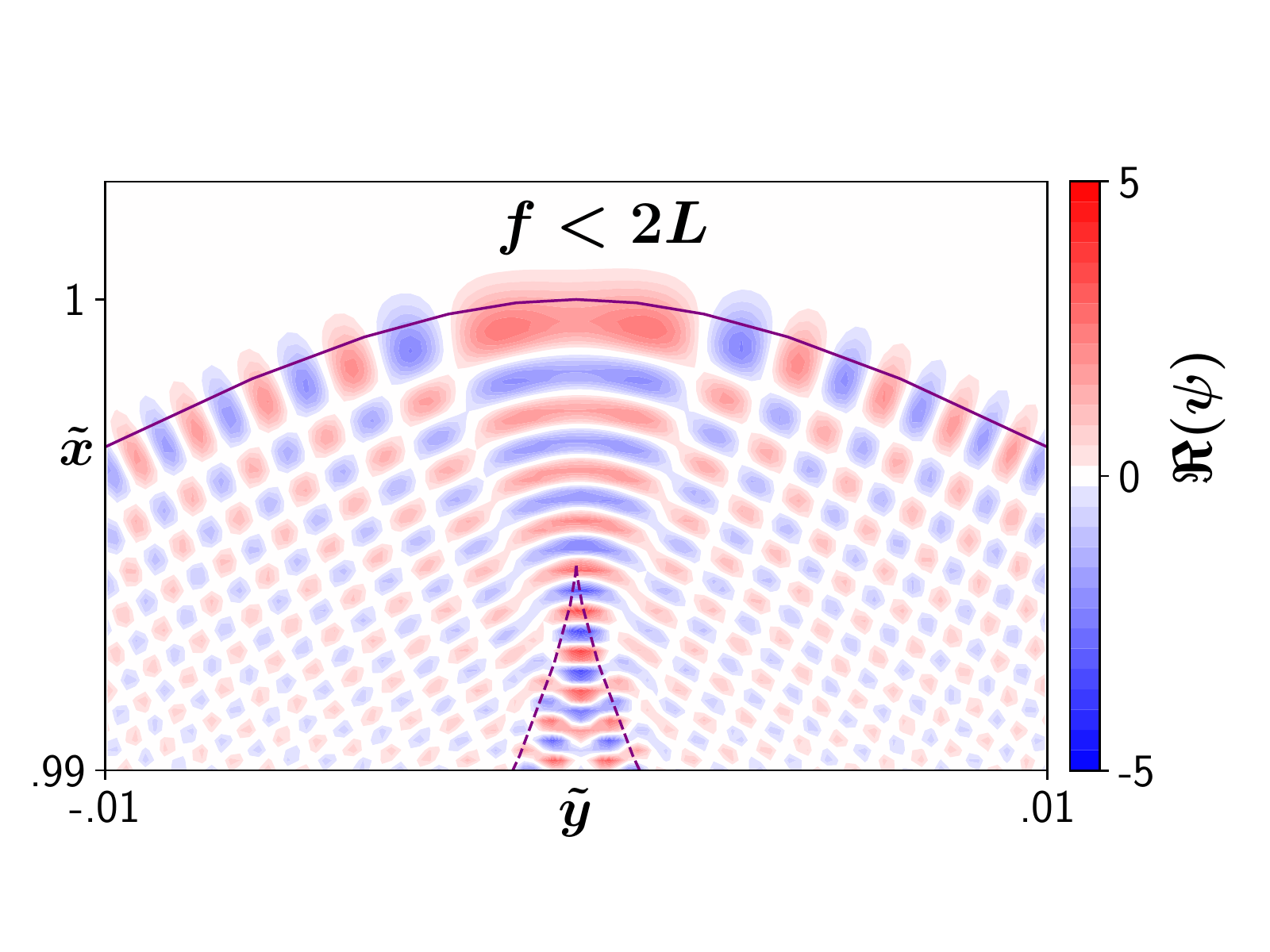}

    \vspace{1mm}
	\includegraphics[width=0.8\linewidth,trim={4mm 14mm 5mm 21mm},clip]{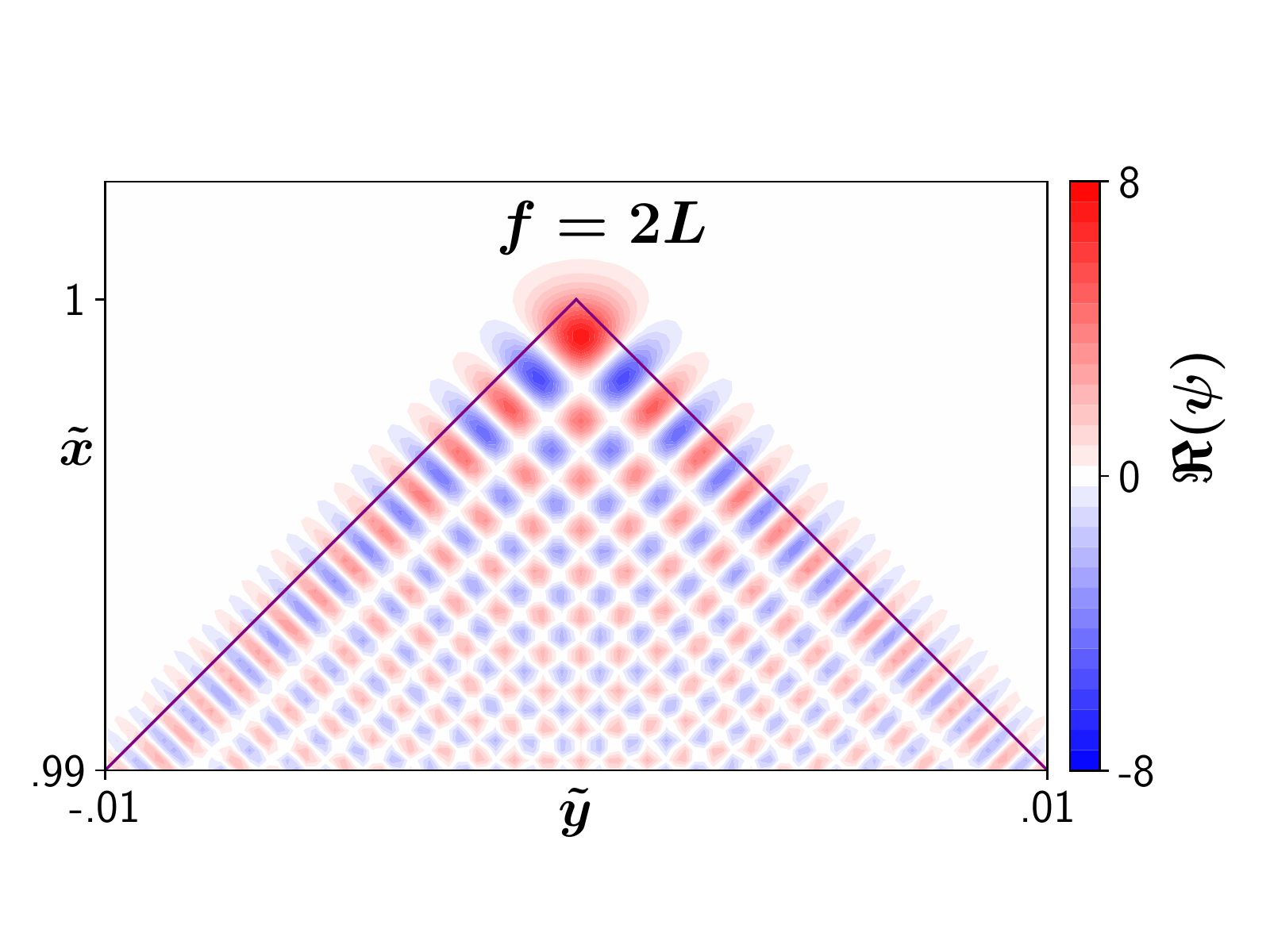}
	
	\vspace{1mm}
	\includegraphics[width=0.8\linewidth,trim={4mm 14mm 5mm 21mm},clip]{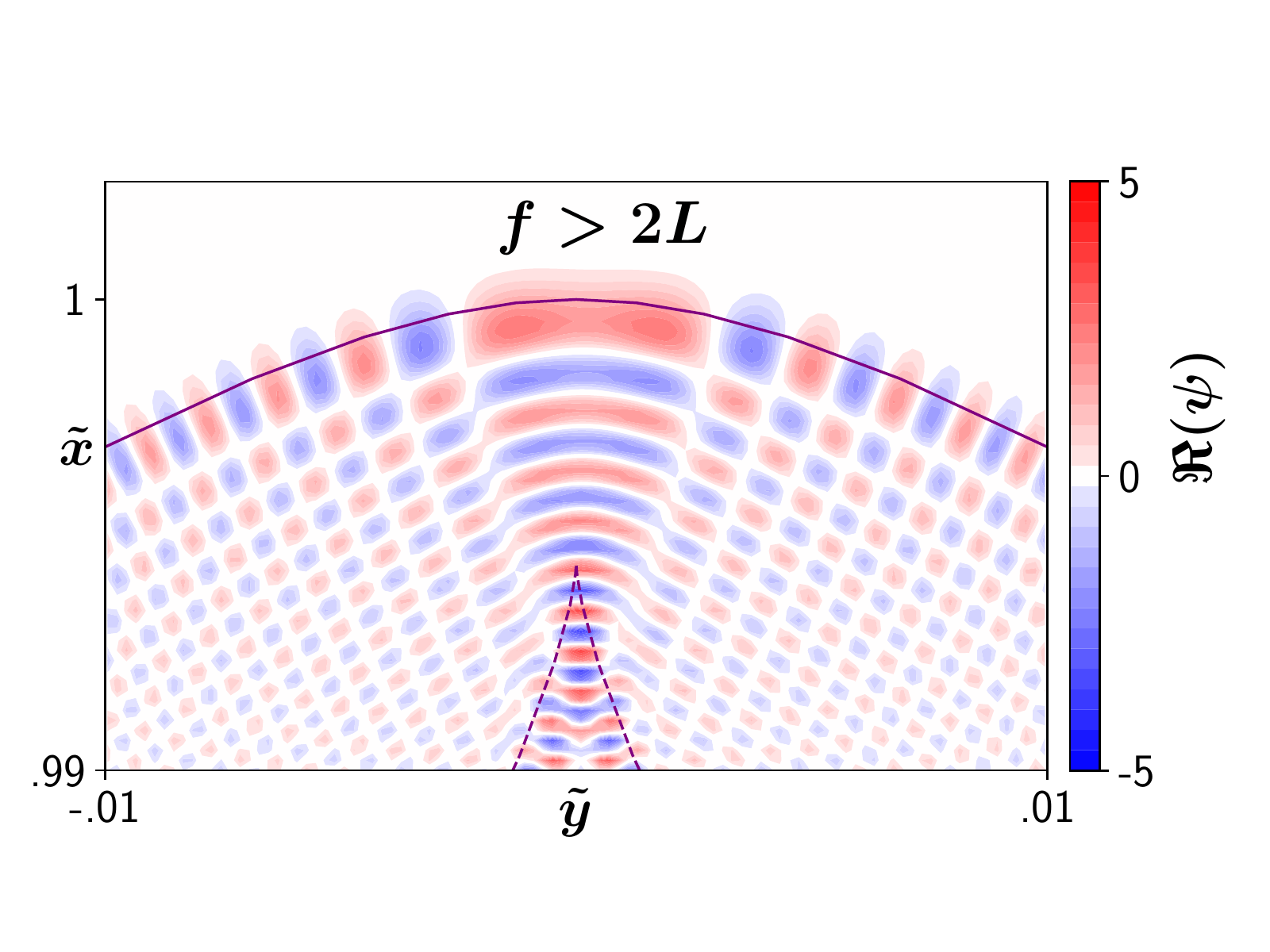}
	\caption{The real part of the hyperbolic umbilic function, \ie the solution \eq{eq:gaussSOL} with $\normONE = 1$, $\nu = 0$, $\waist = 0$, and $\theta = 0$ that describes a focused plane wave propagating in a linear plasma density profile at normal incidence. The purple lines mark the caustics given by \Eqs{eq:critCAUSTIC} or \eq{eq:noncritCAUSTIC}. For oblique incidence, the field pattern simply shifts laterally in $y$ by the amount $2 \airyLEN \sin^2 \theta \tan \theta + \focal \tan \theta \sec \theta$, where the first term is the Goos--H\"anchen shift and the second term corresponds to the shift in the launching point of the field component with transverse wavenumber equal to zero [\Eq{eq:ky0}]. Note that the spatial coordinates are normalized by $\airyLEN$, and the distance between $\tilde{x} = 1$ and $\tilde{x} = 0.99$ constitutes $20$ Airy skindepths, \ie $\airyLEN/\airySKIN = 2000$.}
	\label{fig:HUmbilicFIELD}
\end{figure}

\noindent Importantly, one should keep in mind that the paraxial condition \eq{eq:asympAIRYineq} on the initial conditions requires that the complex focal length be sufficiently large:
\begin{equation}
	\hspace{-2mm}
    | \tilde{\compFOCAL} | \gg \frac{2\pi}{\ell^{3/2}} 
	\max\left[
		1, 
		4 \tan^2 \theta, 
		\frac{\cos^2 \theta}{\left(\sqrt{1 - \ell^{-1}} - \sin \theta\right)^2 }
	\right]
	,
	\label{eq:focalINEQ}
\end{equation}

\noindent where $\ell \doteq \airyLEN/\airySKIN$ and $\tilde{\compFOCAL} \doteq \compFOCAL/\airyLEN$. One also requires the necessary condition \eq{eq:necessary} to be satisfied. The solution \eq{eq:gaussSOL} for $\normONE = 1$, $\nu = 0$, $\waist = 0$, and $\theta = 0$ is shown in \Fig{fig:HUmbilicFIELD}. (Choosing $\nu \neq 0$ and $\waist \neq 0$ shifts the caustic into the complex domain; see \Ref{Colaitis19a} for a detailed discussion of the analogous phenomenon for the Airy function.) 


\begin{figure}
	\includegraphics[width=0.8\linewidth,trim={4mm 14mm 5mm 21mm},clip]{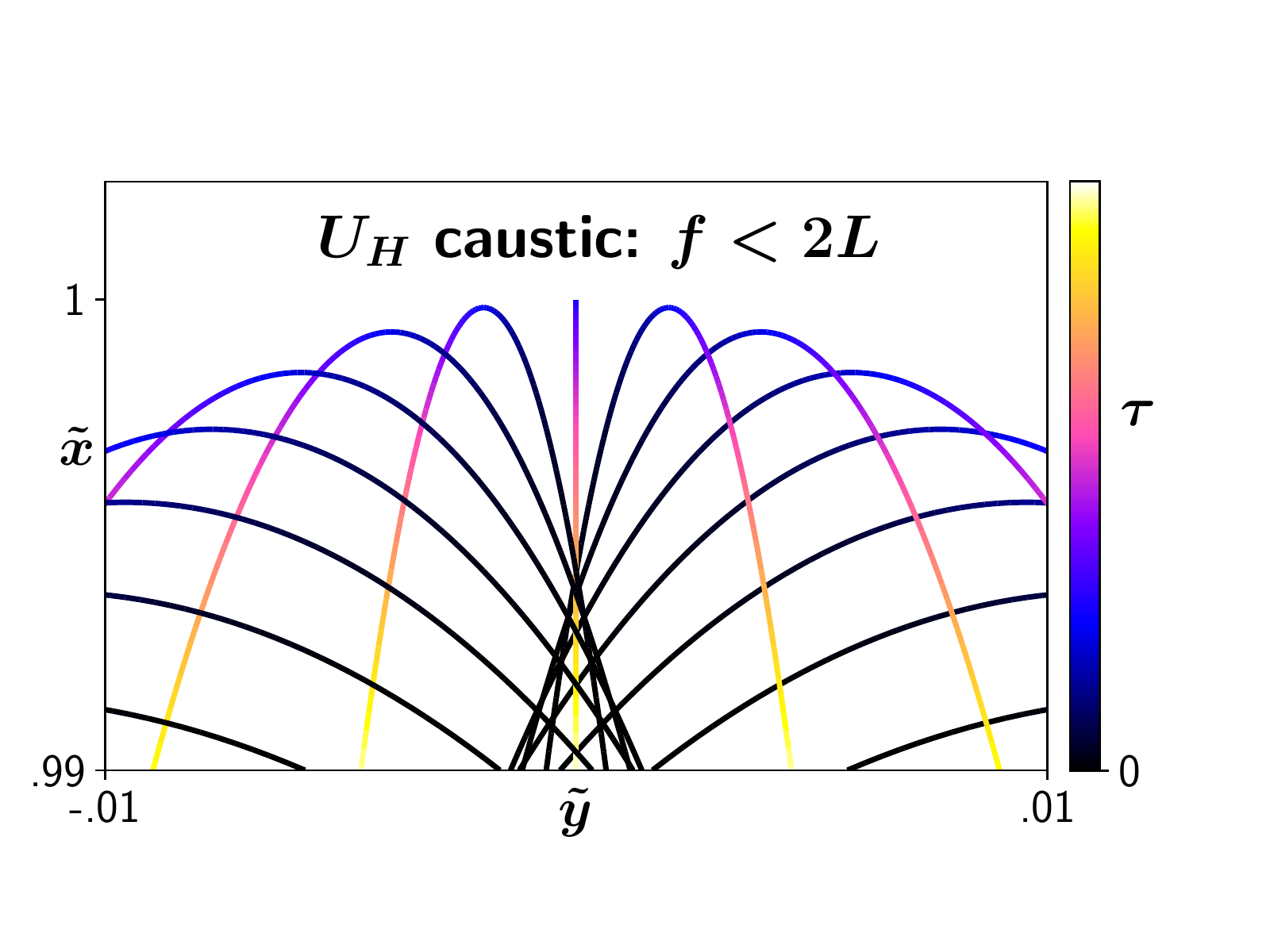}
	
	\vspace{1mm}
	\includegraphics[width=0.8\linewidth,trim={4mm 14mm 5mm 21mm},clip]{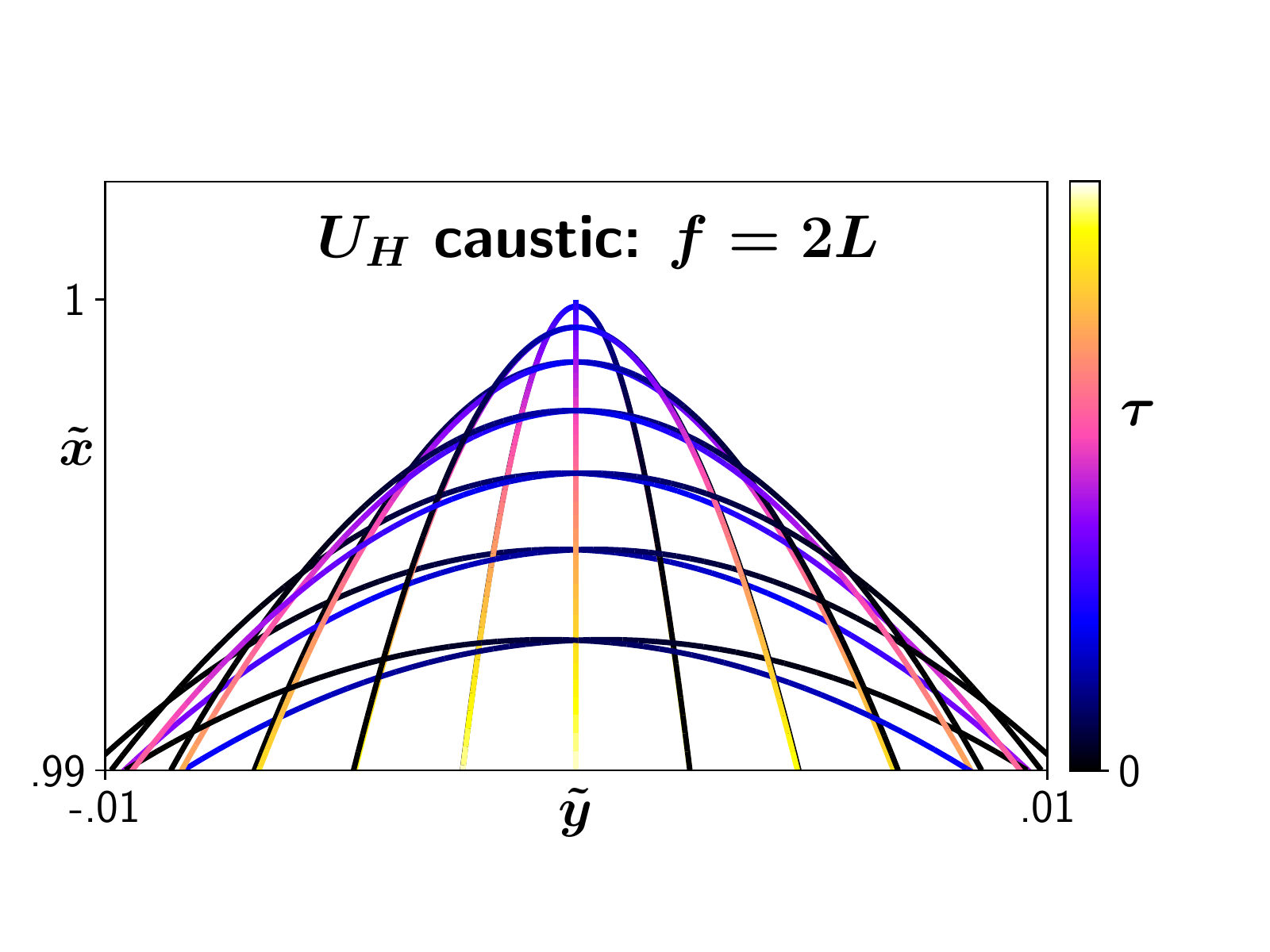}
	
	\vspace{1mm}
	\includegraphics[width=0.8\linewidth,trim={4mm 14mm 5mm 21mm},clip]{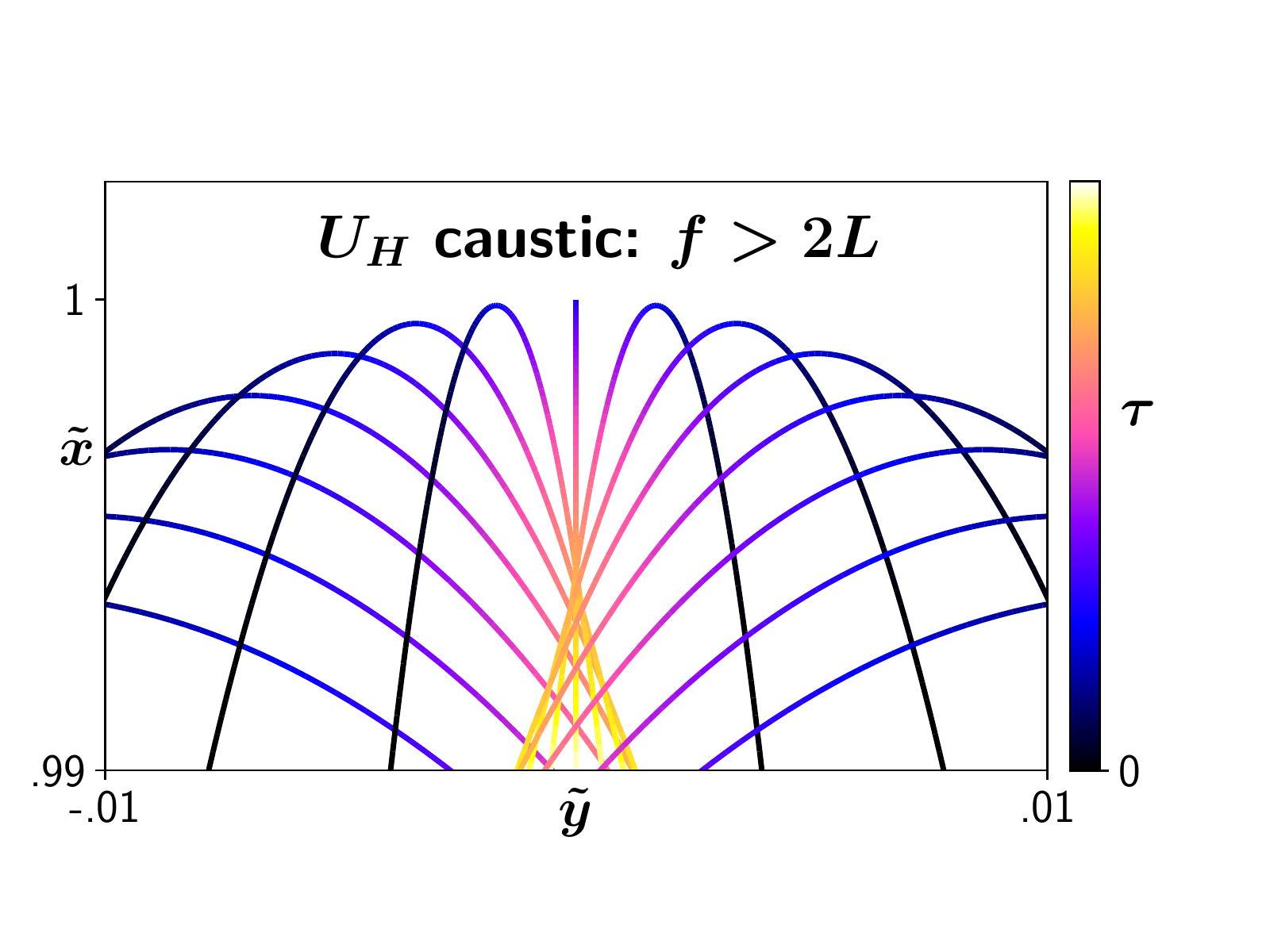}
	\caption{Underlying ray trajectories (\App{app:rays}) of the hyperbolic umbilic function shown in \Fig{fig:HUmbilicFIELD}. Here, $\tau$ denotes the time along a ray: rays are launched at $\tau = 0$ and propagate with increasing $\tau$. Hence, rays focus before reflecting off the critical density layer at $x = \airyLEN$ when $\focal < 2 \airyLEN$, and focus after reflecting off the critical density layer when $\focal > 2 \airyLEN$. Moreover, the density gradient causes the focal spot to become aberrated as a cusp.}
	\label{fig:HUmbilicRAYS}
\end{figure}


Hence, we see that $\hUMB$ can be intuitively understood as the field pattern that results when a focused plane wave of infinite extent encounters a simple turning point~\cite{Orlov80,Kravtsov83}. This intuitive understanding is aided by considering a ray-based description of the wavefield propagation (\App{app:rays}). Figure \ref{fig:HUmbilicRAYS} shows ray trajectories that underlie $\hUMB$ for three cases: \textbf{(i)} $\focal < 2 \airyLEN$, \textbf{(ii)} $\focal = 2 \airyLEN$, and \textbf{(iii)} $\focal > 2 \airyLEN$ at $\theta = 0$. Generally speaking, the focused rays enter the plasma and refract off the density profile according to their angle of incidence, which causes the focal point to become aberrated. When $\focal \neq 2 \airyLEN$, the rays either focus before reflecting off the high-density region or after, in accordance with the sign of $\focal - 2 \airyLEN$; for the special case of $\focal = 2 \airyLEN$ there is actually no aberrated focus, corresponding to the critical point of the hyperbolic umbilic function. The factor of two in the critical focal length is due to the enhanced gradient-index focusing~\cite{Gomez02} of the inhomogeneous plasma density profile - the incident beam must be focused nominally beyond the cutoff to compensate. Also note that the ray equations are unable to describe the Goos--H\"anchen and focal shifts that occur at finite $\theta$, which is typical for such phenomena~\cite{Bliokh13}.

The caustic surfaces where local intensity maxima occur are given by \Eqs{eq:causticTFOLD}-\eq{eq:causticTSEG}, which read in the normalized $(x,y)$ variables (\App{app:rays}) for the critical case $\tilde{\focal} = 2 \cos(2\theta)\cos \theta$ as
\begin{equation}
	\tilde{x} \le 1
	, \quad
	\tilde{y} = \sin(2\theta) \pm (\tilde{x} - 1)
	,
    \label{eq:critCAUSTIC}
\end{equation}

\noindent and for the general case $\tilde{\focal} \neq 2 \cos(2\theta)\cos \theta$ by the parametric curves%
\begin{subequations}%
    \label{eq:noncritCAUSTIC}
	\begin{align}
		\tilde{x}^{\text{(fold)}} &= 
		1 - \Delta \cosh(s)
		\frac{\cosh(s) - 1}{2}
		, \\
		\tilde{y}^{\text{(fold)}} &= 
		\sin(2 \theta)
		+ 2 \sqrt{\Delta}\sin \theta
		+ \Delta \sinh(s)
		\frac{\cosh(s) + 1}{2}
		, \\
		\tilde{x}^{\text{(cusp)}} &= 
		1 - \Delta \cosh(s)
		\frac{\cosh(s) + 1}{2}
		, \\
		\tilde{y}^{\text{(fold)}} &= 
		\sin(2 \theta)
		+ 2 \sqrt{\Delta}\sin \theta
		+ \Delta \sinh(s) \frac{\cosh(s) - 1}{2} 
		,
	\end{align}
\end{subequations}

\noindent where $s \in (-\infty, \infty)$ is the curve parameterization and the fold-cusp separation distance $\Delta$ is given by
\begin{equation}
	\Delta = \left[
		\frac{\tilde{f} - 2\cos(2\theta)\cos \theta}{2 \cos^2 \theta} 
	\right]^2
	.
\end{equation}

\noindent We see the caustic of $\hUMB$ typically consists of two parts: a parabolic-like fold line that constitutes the locus of turning points for the entire wavefield, and a semicubical-like cusp line that corresponds to a focal point aberrated by the plasma density gradient. The hyperbolic umbilic caustic stabilizes this fold-cusp network to perturbations in the problem setup, \eg having a plasma density that deviates from the linear profile assumed here, or having a finite injection angle (the stability of which we have shown explicitly here). 

When the incident wavefield is focused far from the critical density, the two caustic curves separate and the behavior near the turning point is well-described by an Airy function whose level sets are approximately parabolic [\Eq{eq:UHairy}]. The intensity swelling factor would then be given by the usual formula \eq{eq:swellAIRY}. More generally, though, the two caustic curves influence each other to yield a field structure that is more sharply peaked than either the Airy or Pearcey function would predict alone~\cite{Berry80b,Kravtsov83,Poston96, Olver10a}. Indeed, since the peak value of $\hUMB$ is approximately equal to $8$ in the critically focused case (cf.~\Fig{fig:HUmbilicFIELD}), one can estimate the swelling factor for the intensity of \Eq{eq:gaussSOL} as
\begin{equation}
	\left|
		\frac{\psi_\text{max}}{\psi_\text{in}}
	\right|^2
	\sim
	32
	\focal
	\sqrt[3]{ \frac{12}{\airyLEN \lambda^2 \pi^4} } \,
	\cos \theta
	.
	\label{eq:swellHUmb}
\end{equation}

\begin{figure}
    \includegraphics[width=0.85\linewidth,trim={3mm 6mm 2mm 3mm},clip]{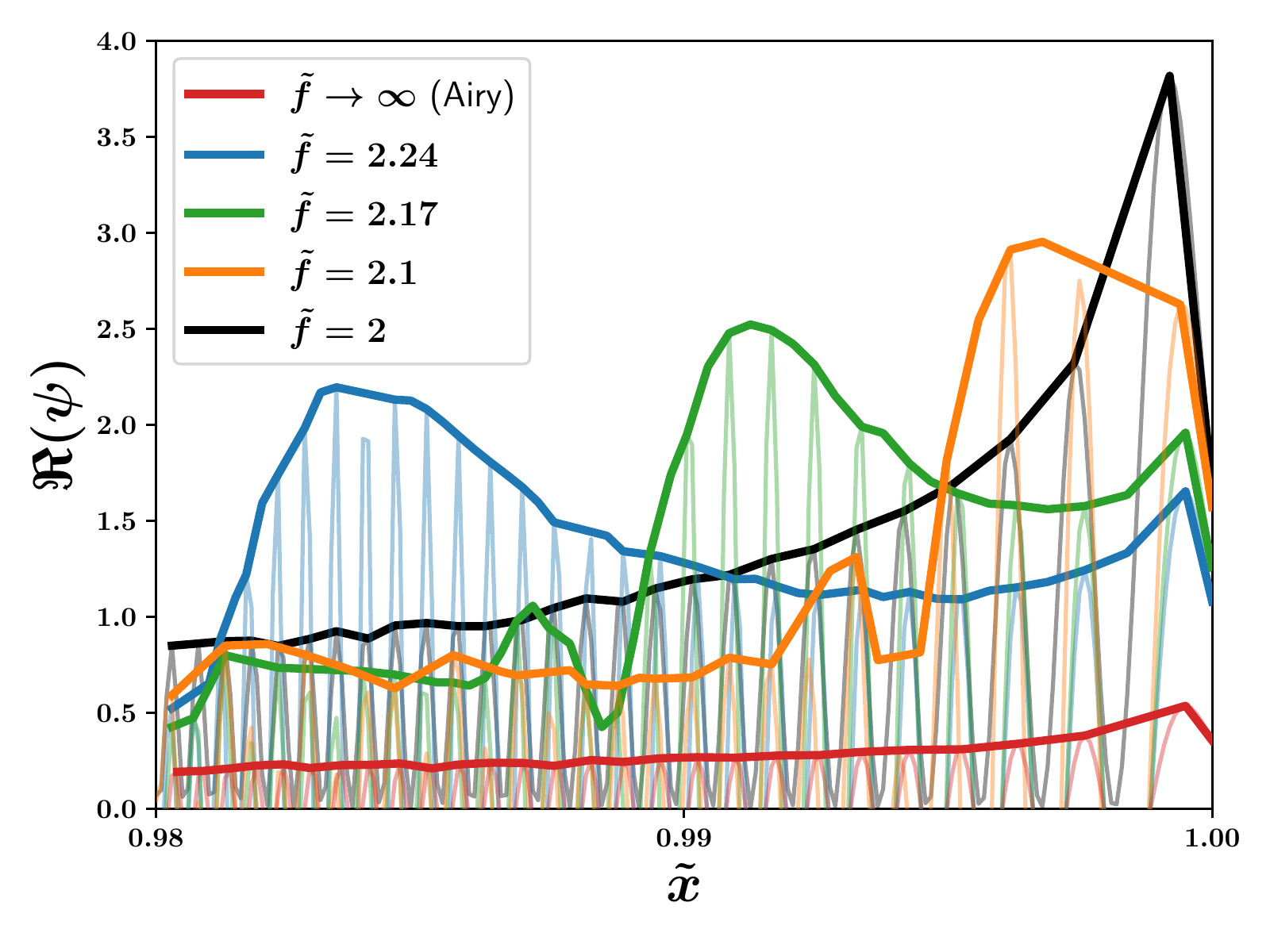}
    \caption{Lineouts of the solution for $\psi$ given by \Eq{eq:gaussSOL} along the symmetry axis $y = 0$ for various values of $\tilde{f}$. All cases have the same initial field amplitude at $\tilde{x} = 0$ given by $E_0 = \sqrt{i \pi} \exp(-i \frac{2}{3} \ell^{3/2})/2 \pi \ell^{1/4}$; the visible differences in the field amplitude at $\tilde{x} = 0.98$ are due to the different swelling factors. The solid color lines shown the field envelope (given as the locus of all local maxima) for a given value of $\tilde{f}$, while the corresponding field oscillations are shown in a lighter shade. Clearly, the swelling of the field increases as the critical focusing condition $\tilde{f} = 2$ is approached. Note also that $\tilde{f} \to \infty$ recovers the standard Airy function solution (red), whose swelling factor is paltry in comparison.}
    \label{fig:swell}
\end{figure}

\noindent Again, the power-law scaling of the swelling factor \eq{eq:swellHUmb} with respect to $1/\lambda$ ($2/3$) is equal to twice the singularity index of the hyperbolic umbilic catastrophe function~\cite{Berry80b,Olver10a} ($1/3$) by definition. The enhanced swelling of $\hUMB$ compared to $\airyA$ is demonstrated in \Fig{fig:swell}, which shows lineouts along the axis $y = 0$ for $\hUMB$ at various values of $\tilde{f}$, including at $\tilde{f} \to \infty$ when $\hUMB$ reduces to $\airyA$ [\ie \Eq{eq:gaussSOL} reduces to \Eq{eq:obliqueAIRY}].

\begin{figure}
	\includegraphics[width=0.85\linewidth,trim = 3mm 4mm 3mm 3mm, clip]{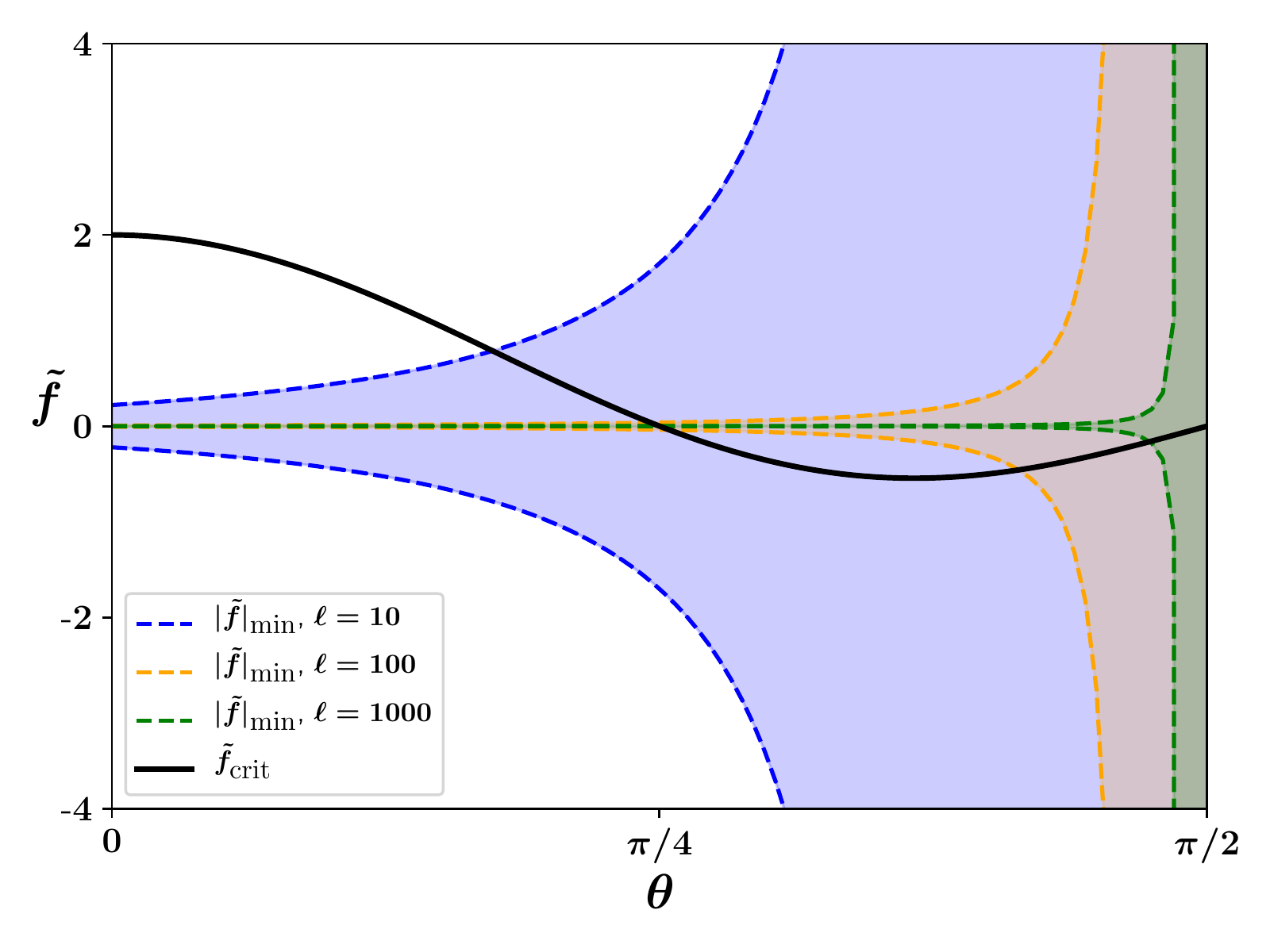}
	\caption{Comparison for various values of the normalized density scale length $\ell$ between the critical focal length $\tilde{f}_\textrm{crit}$ (black solid curve) and the minimum value $\tilde{f}_\text{min}$ given by \Eq{eq:focalINEQ} that is required for the validity of \Eq{eq:gaussSOL}. At a given $\ell$, the critical focal length, and thus the most singular behavior of $\hUMB$ (\Fig{fig:HUmbilicFIELD}) with enhanced swelling given by \Eq{eq:swellHUmb}, is only achievable for the range of obliquity angles $\theta$ for which the black line lies outside the region shaded by the respective color. Note that the focal length is normalized by the density scale length. For $\lambda = 351$~nm, the absolute lengthscales $\airyLEN$ shown in the plot are given respectively as $0.002$~mm (blue), $0.05$~mm (orange), and $1.7$~mm (green).}
	\label{fig:focalCOMP}
\end{figure}

That said, however, the paraxial constraint \eq{eq:focalINEQ} on the focal length means that one is not always able to realize the full morphology of $\hUMB$ within our model, depending on the injection angle and the density lengthscale. A comparison between the critical focal length
\begin{equation}
	\tilde{f}_\textrm{crit} = 2 \cos(2 \theta) \cos\theta
\end{equation}

\noindent that would create the most singular behavior of $\hUMB$ and the minimum focal length set by \Eq{eq:focalINEQ} is shown in \Fig{fig:focalCOMP} for various values of $\ell$. Immediately, one makes the curious observation that the critical focal length crosses zero at $\theta = \pi/4$ and becomes negative. This suggests that at large oblique angles the density gradient introduces such strong focal aberrations that one must launch a defocused (expanding) beam to obtain the critical behavior. Additionally, from the figure one sees that our model is only applicable over a small range of $\theta$ and $\tilde{f}$ when the normalized density lengthscale is relatively short. As the lengthscale gets longer, the region of validity increases, although the zero crossing at $\theta = \pi/2$ always remains outside this region. (Note that for the NIF laser~\cite{Spaeth17} with $\lambda = 351$~nm, the normalized lengthscales $\ell = 10$, $100$, and $1000$ shown \Fig{fig:focalCOMP} correspond to absolute lengthscales of $0.002$~mm, $0.05$~mm, and $1.7$~mm respectively.)

It is interesting to then consider how our model can be applied to present experiments. In particular, real experiments have a density profile that evolves in time. Consider the isothermal expansion of a laser-ablated hohlraum plasma for example. In such a plasma, the density lengthscale will increase as $\airyLEN \sim C_s t$ with 
\begin{equation}
    C_s \doteq
    \sqrt{\frac{Z T_e}{m_i}}
\end{equation}

\noindent being the sound speed, and correspondingly, our model's region of validity will steadily expand in time. Using the necessary condition \eq{eq:necessary} as means of a simple estimate, the NIF outer versus inner beams (which respectively make angles $\theta \sim 40^\circ$ and $\theta \sim 60^\circ$ with the hohlraum wall normal~\cite{Spaeth17}) can then be described after time $t \gtrsim 1.8$~ps and $t \gtrsim 6.4$~ps respectively. (Note that we have taken $T_e \sim 0.5$~keV, $Z \sim 20$, and $m_i = m_{Au}$ as typical early-time parameters for the ablating plasma.) This necessary delay time before our model can be applied to NIF-like parameters is negligible compared to the nanosecond timescales of the experiments. Furthermore, the steadily increasing Goos-H\"anchen shift of the reflection point (equal to $2 \airyLEN \tan \theta \sin^2 \theta$) as time progresses suggests that ray-tracing calculations (which do not contain this shift) may become increasingly inaccurate at later times. This observation is particularly relevant because of the recent interest in characterizing glint losses in hohlraums~\cite{Lemos22}; if the reflection geometry is not specular but instead has angle-dependent shifts, the interpretation for which lasers are responsible for which glint signals may change (although the increased collisional absorption that also occurs at large $\airyLEN$ may dominate this effect in certain parameter regimes~\cite{Kruer03}).

As the density evolves, the focusing of the incident wavefield will change even if the focal length of the lens remains the same. This is because increasing $\airyLEN$ drives $\tilde{f}$ towards zero. A qualitative understanding of this effect can be readily obtained by viewing \Fig{fig:focalCOMP}: when $\theta \le \pi/4$ a wave with initial $\tilde{f} < 0$ will get more focused but never critically focused since it cannot cross $\tilde{f} = 0$, while a wave with $\tilde{f} > \tilde{f}_\text{crit}$ will become critically focused and then defocus; conversely, when $\theta > \pi/4$ a wave with $\tilde{f} \ge 0$ will become more focused but never critically focused, while a wave with $\tilde{f} < \tilde{f}_\text{crit}$ will pass through critically focusing on its way to becoming defocused. This generic behavior should be observable in other wave applications too, for example, electron cyclotron resonance heating on spherical tokamaks during the density rampup phase~\cite{Lopez18a}. Since ray-tracing codes are often used to optimize such applications, this observation means that advanced ray-tracing techniques such as etalon integrals~\cite{Colaitis19a, Colaitis19b, Colaitis21} or metaplectic geometrical optics~\cite{Lopez20,Lopez21a,Lopez22,Lopez22t} are needed to enable the accurate computation of the entire unfolding of $\hUMB$.


\subsection{Finite-aperature effects}
\label{sec:aper}

Now let us consider how the presence of an aperture might modify the results thus far obtained. This is accomplished by letting%
\footnote{Note that our notion of an aperture is different from that used in the detailed study of \Ref{Nye06a}.}%
\begin{equation}
    \psi_\text{in}(0,y)
    = \Psi_\text{in}(0,y) \, \text{rect}\left( \frac{y}{\width}\right)
    ,
\end{equation}

\noindent where $\text{rect}(z)$ denotes the rectangular hat function, which is everywhere zero except when $-1/2 < z < 1/2$ where it equals unity. The shifted FT of $\psi$ is then given by the usual convolution formula
\begin{equation}
    \fourier{\psi}_\text{in}(k_y)
    = \frac{\width}{2\pi}
    \int \dd \kappa_y \,
    \sinc\left(
        \frac{\width \kappa_y}{2}
    \right)
    \fourier{\Psi}_\text{in}(k_y - \kappa_y)
    .
\end{equation}

\noindent In view of \Eq{eq:AIRYsol}, if $\width$ is much larger than the characteristic variations in $\Psi$ and $\hUMB$, then one can take $\width \to \infty$ such that $\psi \approx \Psi$, meaning that the aperture plays no role. Similarly, if $\width$ is much smaller than the characteristic variations, one can take $\width \to 0$ such that $\text{rect}\left( \frac{y}{\width}\right) \approx \width \delta(y)$ and one correspondingly obtains
\begin{widetext}
    \begin{align}
        \psi(x, y)
        &\approx \width \normTWO \, \Psi_\text{in}(0,0)
        \hUMB
        \left(
            \sqrt[3]{3} \frac{x - \compLEN}{\airySKIN}
            ,
            \frac{ 
                y \cos \theta - 2 \compLEN \sin^3 \theta
            }{\sqrt[6]{3} \, \airySKIN \cos \theta}
            ,
            - \frac{\lambda \compLEN}{2\pi \sqrt[3]{3} \, \airySKIN^2 } \frac{\cos 2 \theta}{\cos \theta}
        \right)
        ,
        \label{eq:tinyAPER}
    \end{align}
\end{widetext}

\begin{figure}
    \includegraphics[width=0.8\linewidth,trim={4mm 13mm 5mm 20mm}, clip]{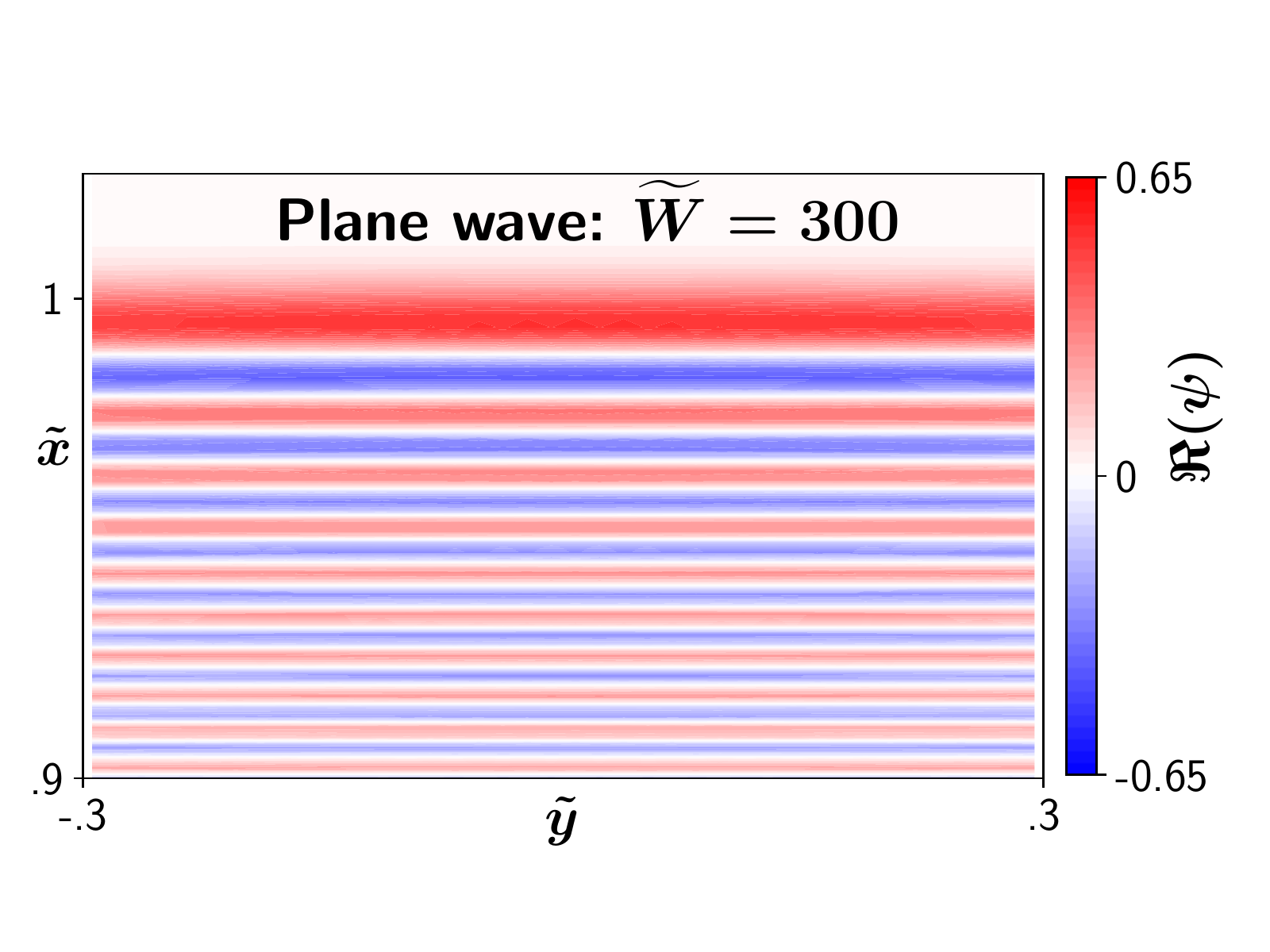}
    
    \includegraphics[width=0.8\linewidth,trim={4mm 13mm 5mm 20mm}, clip]{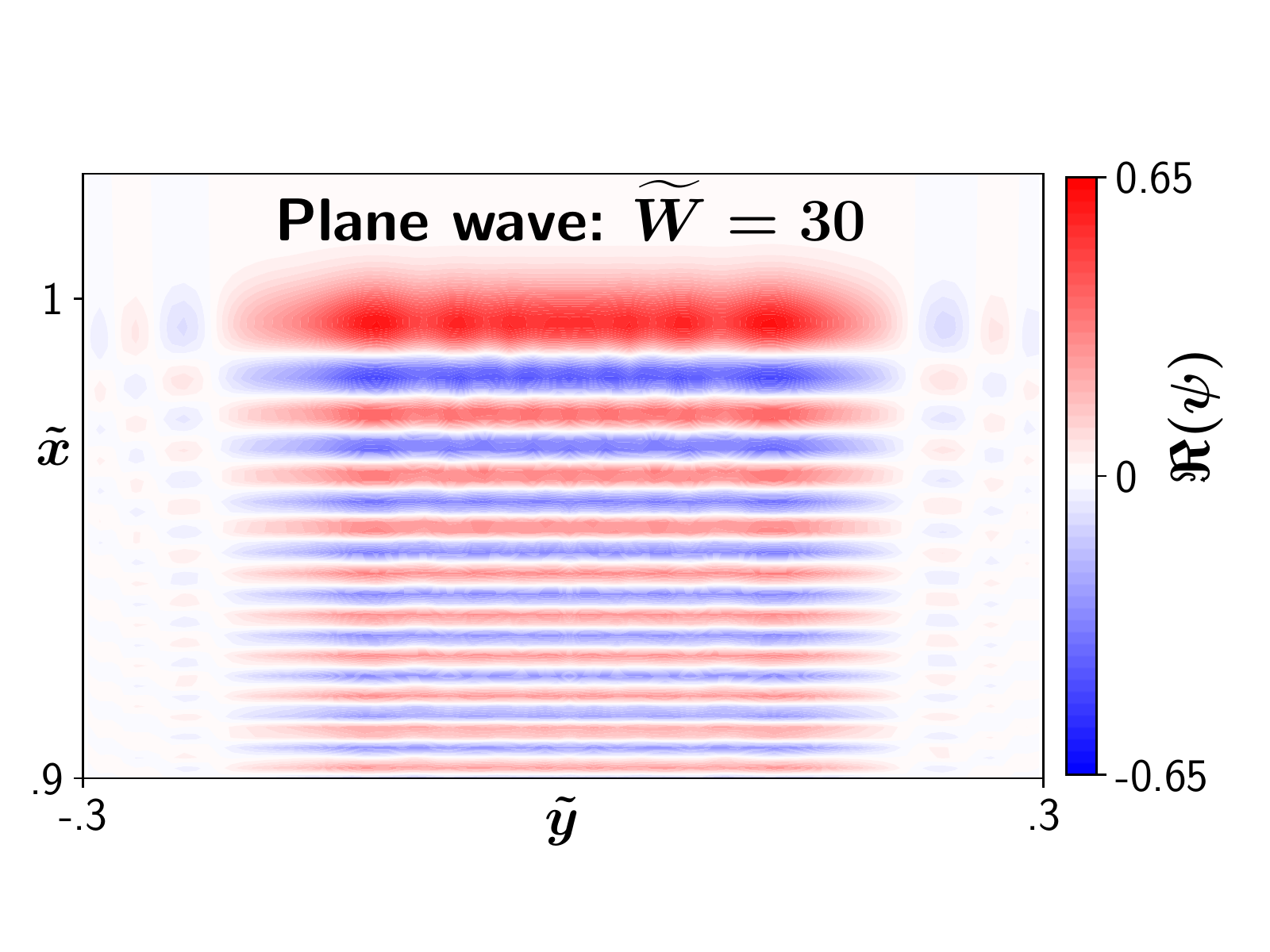}
    
    \includegraphics[width=0.8\linewidth,trim={4mm 13mm 5mm 20mm}, clip]{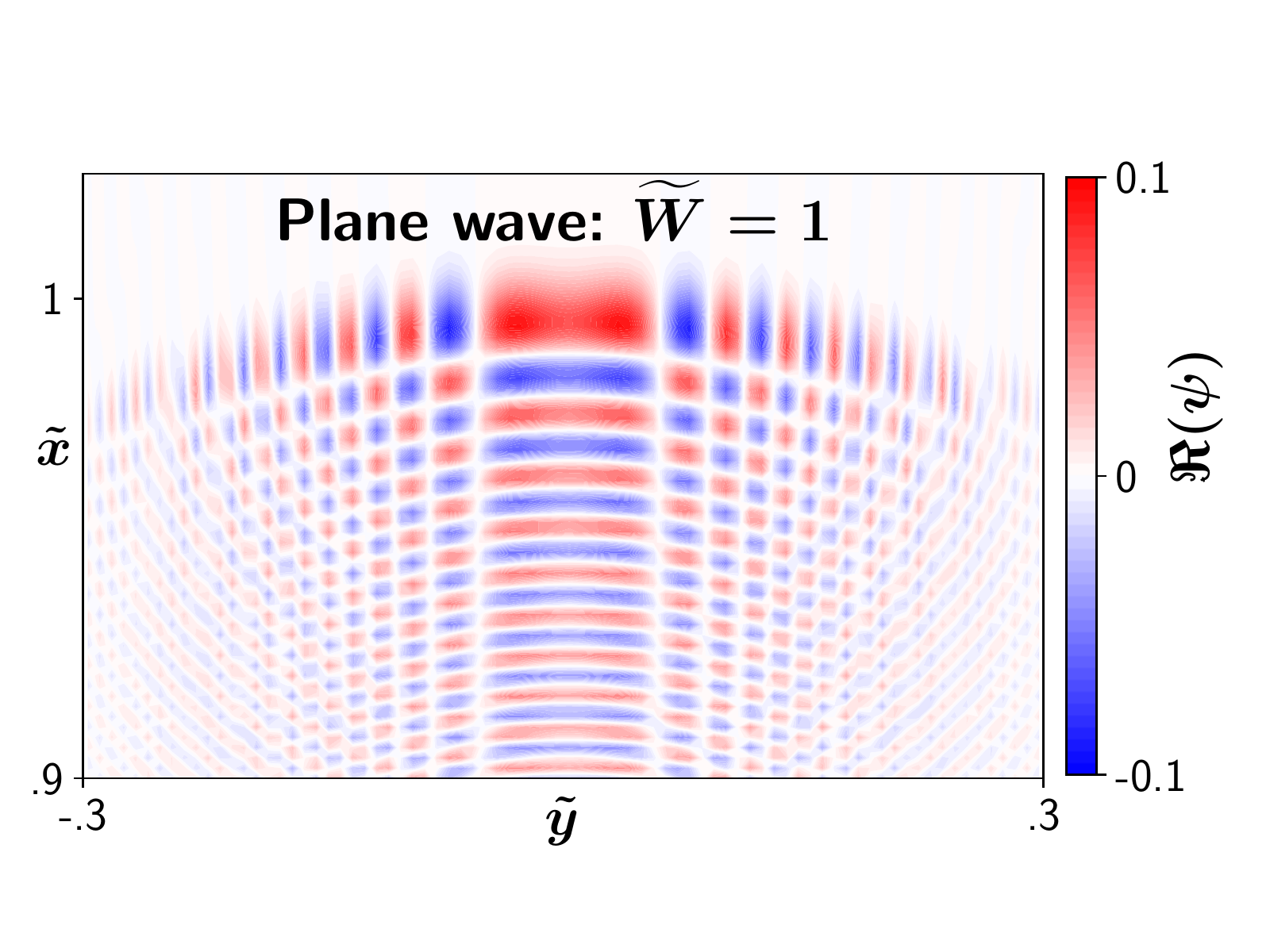}
    \caption{Progression of the diffraction pattern for an apertured plane wave near a turning point located at $\tilde{x} = 1$ as the aperture width $\widetilde{W}$ is varied. The pattern transforms from the standard Airy pattern at large $\widetilde{W}$ to the defocused hyperbolic umbilic function \eq{eq:tinyAPER} at small $\widetilde{W}$ according to \Eq{eq:planeAPER}, which is numerically solved using the method outlined in \App{app:compute}. Here $\airyLEN/\airySKIN = 200$, so the distance between $\tilde{x} = 1$ and $\tilde{x} = 0.9$ constitutes $20$ Airy skindepths.}
    \label{fig:planeAPER}
\end{figure}

\begin{figure}
    \includegraphics[width=0.8\linewidth,trim={4mm 13mm 5mm 20mm}, clip]{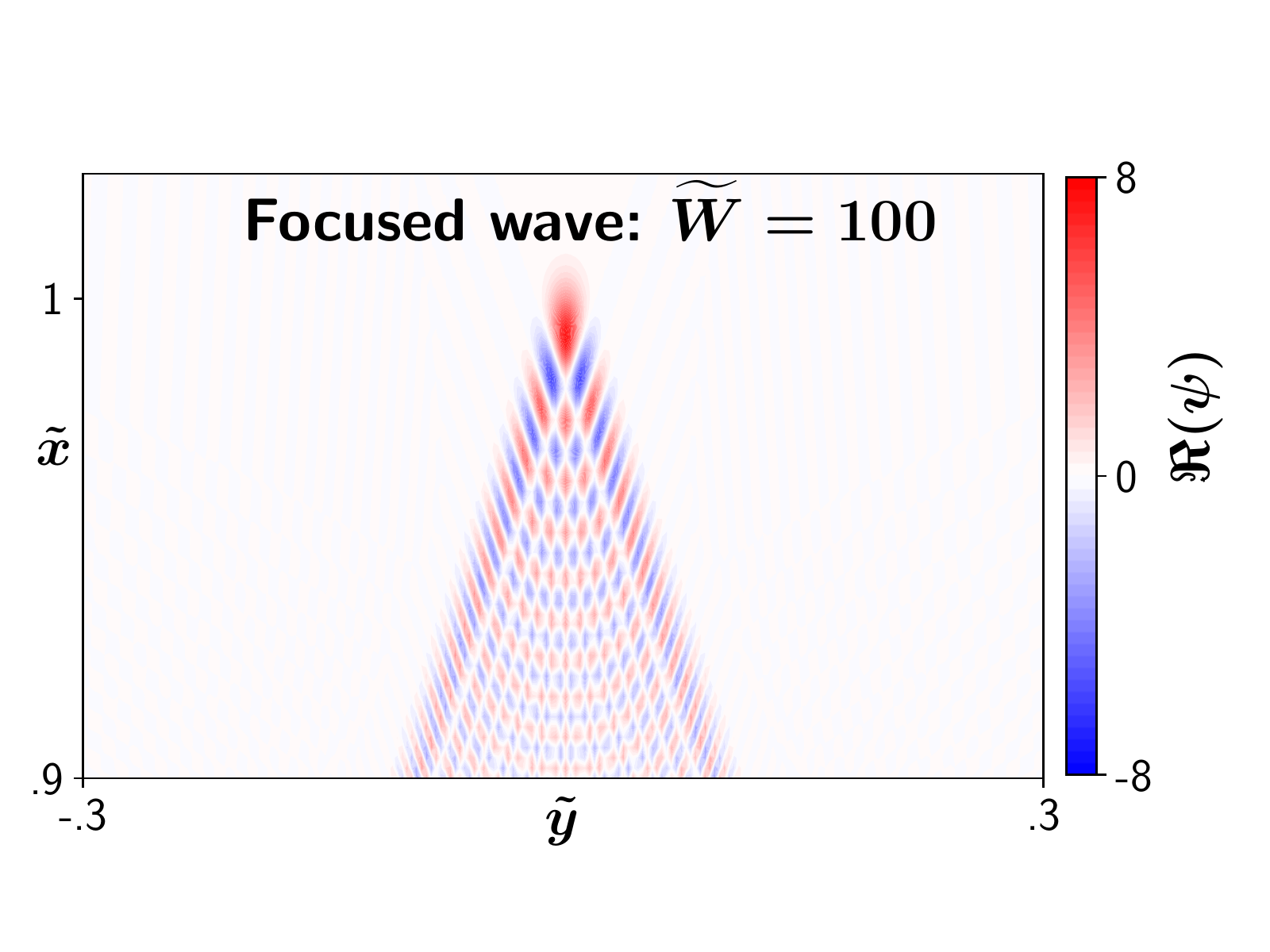}
    
    \includegraphics[width=0.8\linewidth,trim={4mm 13mm 5mm 20mm}, clip]{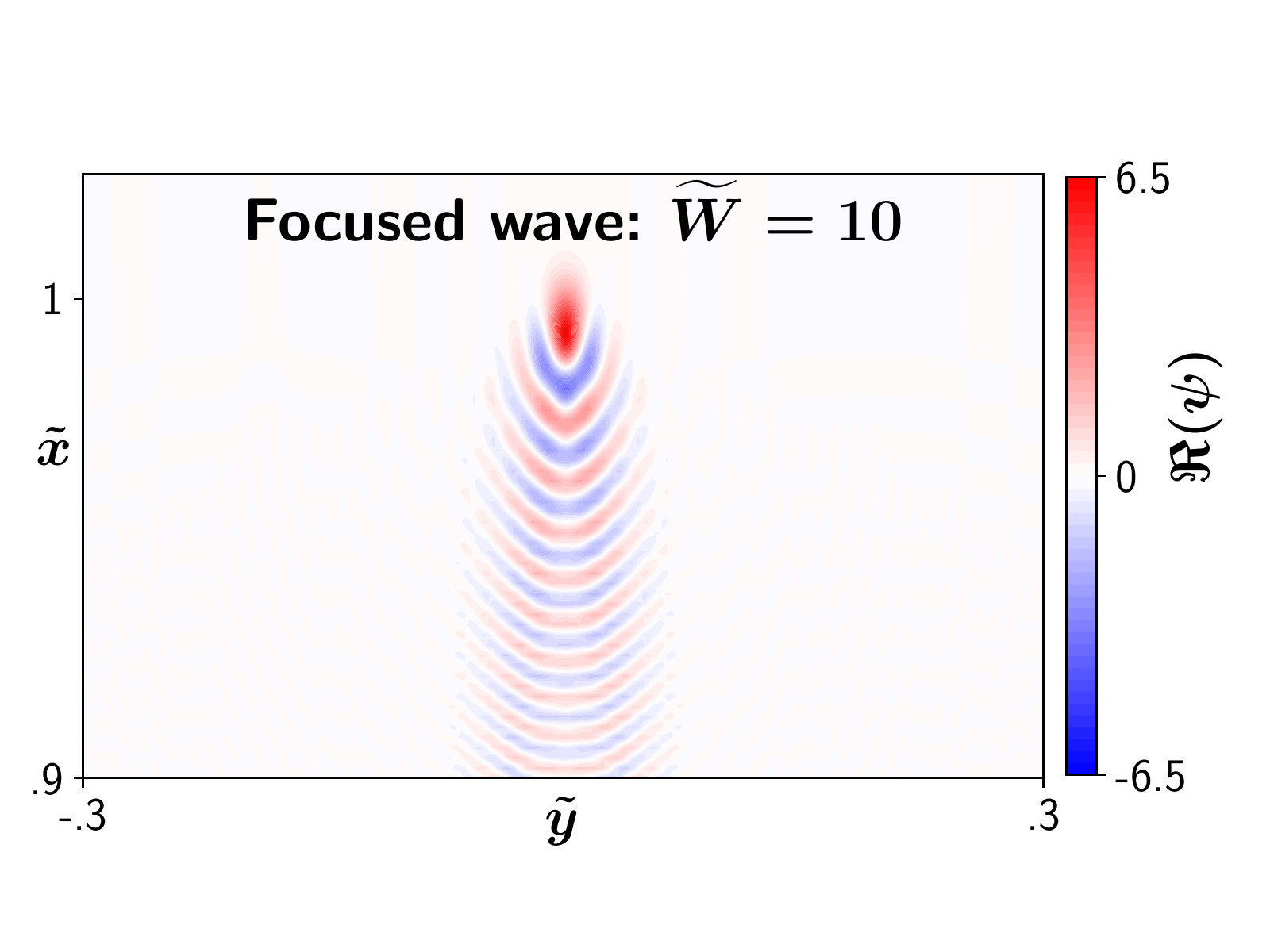}
    
    \includegraphics[width=0.8\linewidth,trim={4mm 13mm 5mm 20mm}, clip]{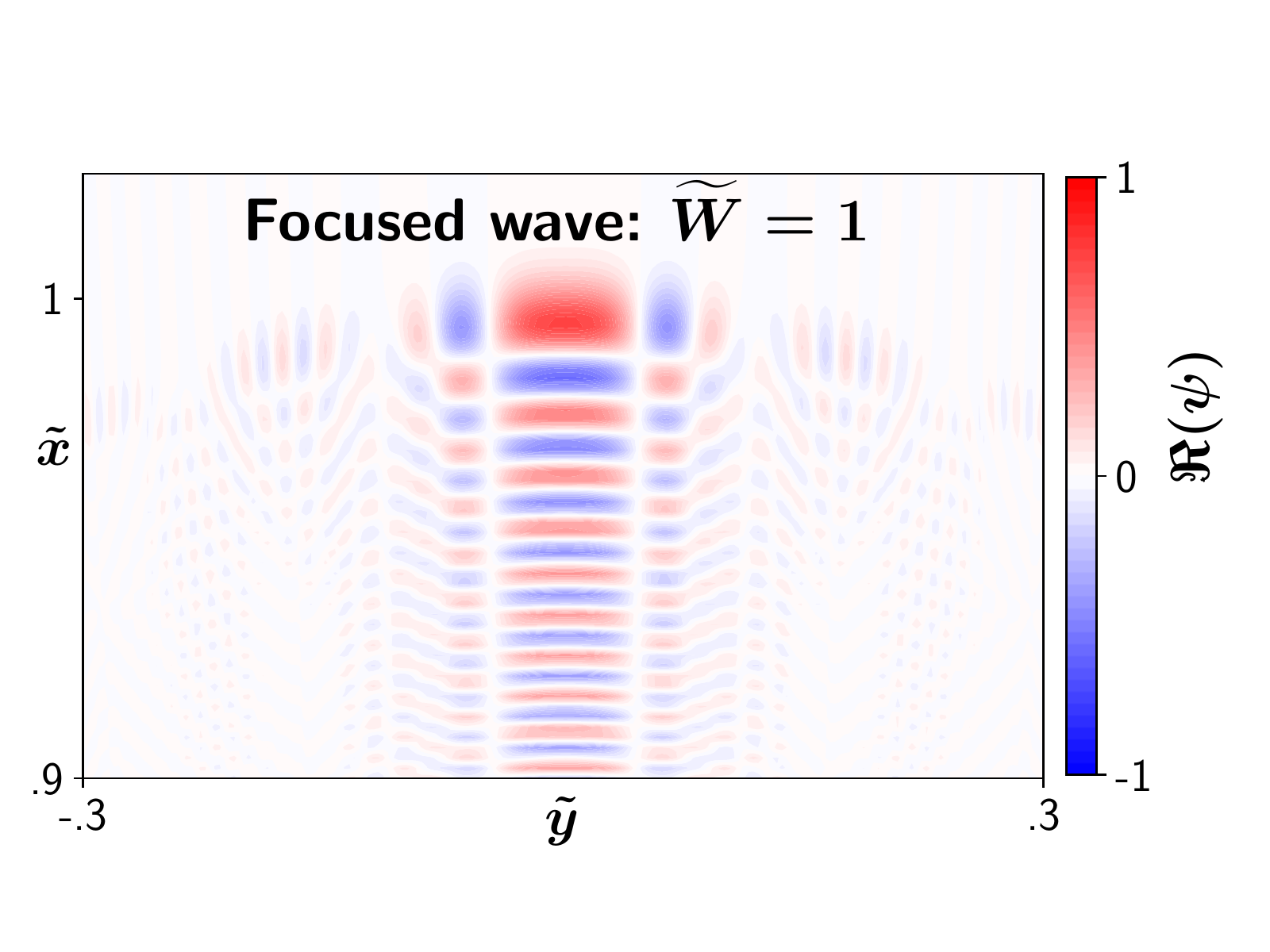}

    \includegraphics[width=0.8\linewidth,trim={4mm 13mm 5mm 20mm}, clip]{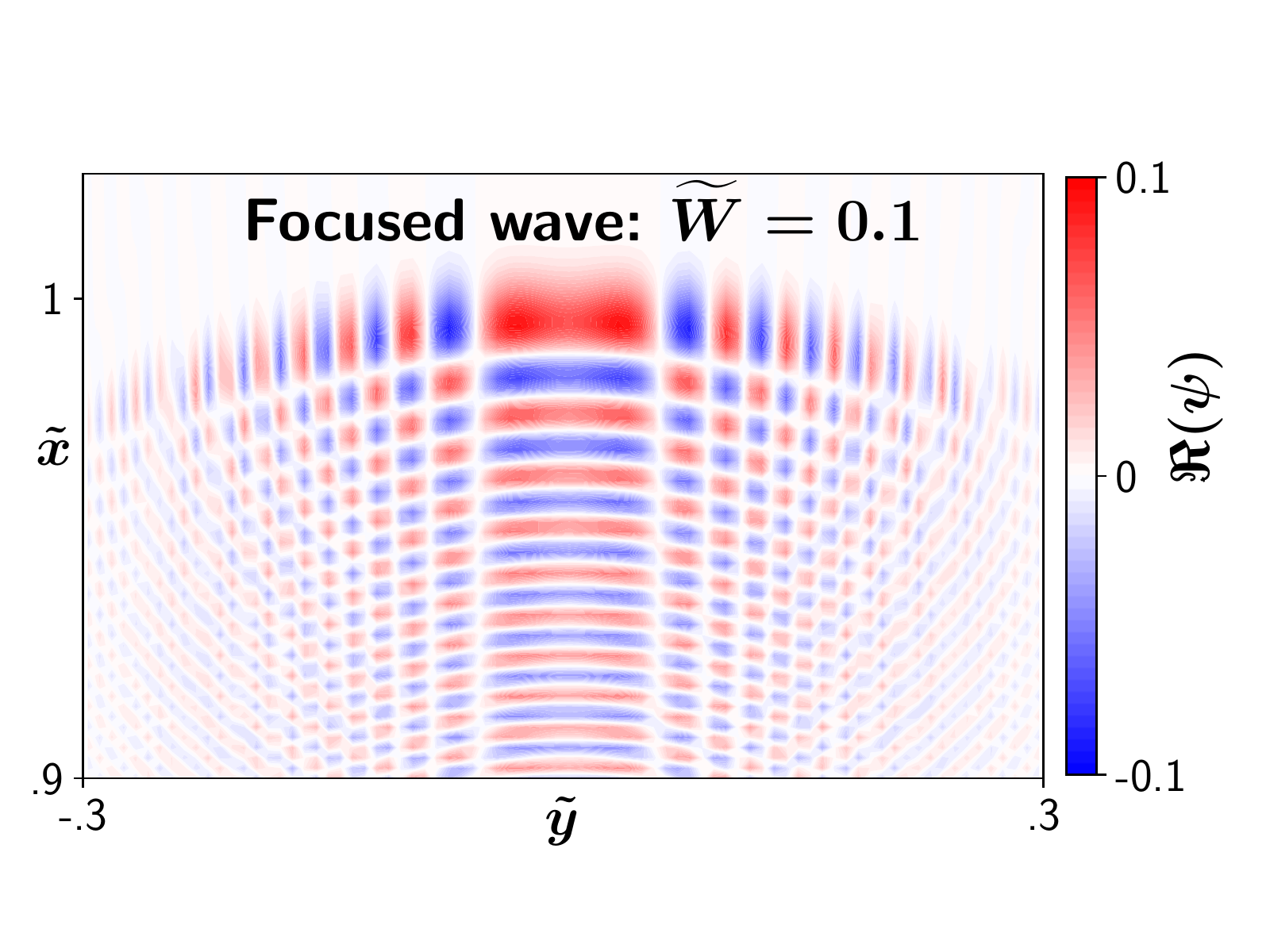}
    \caption{Same as \Fig{fig:planeAPER} but for the apertured focused wave described by \Eq{eq:focusAPER} with critical focusing $\tilde{f} = 2$. The transitions for non-critical $\tilde{f}$ are qualitatively similar.}
    \label{fig:focusAPER}
\end{figure}

\noindent where $\normTWO$ is given by \Eq{eq:originalNORM}. Hence, $\hUMB$ can also be understood as the point-spread function~\cite{Goodman05} for propagation in a linear density gradient, since in this limit the signal that passes through the aperture can be considered a point source. Finite $\width$ will therefore generate a homotopic transformation between the unapertured solution \eq{eq:AIRYsol} and the hyperbolic umbilic solution \eq{eq:tinyAPER}.

For an incident plane wave (\Sec{sec:plane}), the transformation is given explicitly as
\begin{align}
    \psi(x, y)
	&\propto
    \frac{\sqrt[3]{3} \, \widetilde{\width}}{2\pi^2}
	\int \dd u \, \dd v
	\,
    \sinc\left(
        \widetilde{\width}u
    \right)
    \nonumber\\
    &\hspace{4mm}\times
	\hUMBKER
	\left[
		u, v,
		\sqrt[3]{3} \, \ell (\tilde{x} - 1),
		\frac{ \ell }
        {\sqrt[6]{3}} \tilde{y},
		- \frac{ \sqrt{\ell} }{\sqrt[3]{3} }
	\right]
	,
    \label{eq:planeAPER}
\end{align}

\noindent where we have assumed $\theta = \nu = 0$ for simplicity. Also, we have re-introduced the normalized coordinates defined in \Eqs{eq:normCOORD} along with the normalized aperture width
\begin{equation}
    \widetilde{\width} \doteq \frac{\width}{2 \sqrt[6]{3} \, \airySKIN}.
\end{equation}

\noindent One readily verifies that the hyperbolic umbilic solution \eq{eq:tinyAPER} is recovered in the limit $\widetilde{\width} \to 0$ and the Airy solution \eq{eq:obliqueAIRY} is recovered in the limit $\widetilde{\width} \to \infty$. This transition is shown in detail in \Fig{fig:planeAPER}, in which \Eq{eq:planeAPER} is numerically calculated for various values of $\widetilde{\width}$.

Similarly, for an incident focused wave (\Sec{sec:focused}), the transformation is given explicitly as
\begin{align}
    \psi(x, y)
	&\propto
	\int \dd u \, \dd v
	\,
    \frac{\aperFUNC(u)}{2} 
    \nonumber\\
    &\times
	\hUMBKER
	\left[
		u, v,
		\sqrt[3]{3} \ell (\tilde{x} - 1),
		\frac{ \ell}{\sqrt[6]{3}} \tilde{y},
		\frac{\sqrt{\ell}}{\sqrt[3]{3}} \frac{\tilde{\focal} - 2}{2}
	\right]
    ,
    \label{eq:focusAPER}
\end{align}

\noindent where we have taken $\theta = \nu = \waist = 0$, and we have introduced the aperture function
\begin{align}
    \aperFUNC(u) &\doteq
    \erf\left( 
        \sqrt{\frac{i \tilde{\focal}}{2 }} \frac{\sqrt[4]{\ell}}{\sqrt[6]{3}} u
        + \frac{\sqrt{i} }{2} \widetilde{\width}
    \right)
    \nonumber\\
	&- \erf\left( 
        \sqrt{\frac{i \tilde{\focal}}{2 }} \frac{\sqrt[4]{\ell}}{\sqrt[6]{3}} u
        - \frac{\sqrt{i} }{2} \widetilde{\width}
    \right)
    .
\end{align}

\noindent For convenience, we have also altered our normalization convention for the aperture width such that now
\begin{equation}
    \widetilde{\width} = \sqrt{\frac{\pi}{\lambda \focal}} \width
\end{equation}

\noindent (which one also recognizes as simply the Fresnel number of the aperture evaluated at the focal length~\cite{Born99}). Again, one can verify that \Eq{eq:tinyAPER} is recovered from \Eq{eq:focusAPER} in the limit $\widetilde{\width} \to 0$, and that \Eq{eq:gaussSOL} is obtained in the limit $\widetilde{\width} \to \infty$. The transformation between these two limiting cases is depicted in \Fig{fig:focusAPER}, where \Eq{eq:focusAPER} is numerically computed for a sequence of $\widetilde{\width}$ values.

\section{Conclusions}
\label{sec:concl}

In this work a model is proposed to describe the diffraction pattern of a general wavefield incident upon a turning point. Assuming that the incoming field has a bounded Fourier spectrum about the mean angle of incidence, the solution can then be expressed as an integral mapping of the initial wavefield whose kernel is the hyperbolic umbilic function, one of the seven famous functions from catastrophe theory. At normal incidence the integral takes the form of a convolution. It is shown that the traditional Airy solution is subsumed as a special case, and also that the hyperbolic umbilic function is itself a solution when the incident wavefield is Gaussian focused. Also, when the initial field is passed through an aperture, the solution generically transforms from the original aperture-free case to a defocused hyperbolic umbilic function, and explicit examples of this transformation are given for a plane wave and a Gaussian wave.

Due to the ubiquity of focused waves near turning points, the results presented here should have broad applications. In fusion research, these observations may enable the development of more accurate reduced models for lasers interacting with and reflecting off the hohlraum wall. It also lays the foundation for future studies to understand how the reflection physics might be modified by the presence of speckles. Preliminary work~\cite{Lopez21DPP} suggests a speckled laser near a turning point can be described by a random sum of the apertured hyperbolic umbilic functions discussed here, although more analysis is required to confirm this finding and also to explore its consequences on modern ICF experiments. For certain phenomenological speckle models, \eg treating speckled lasers as a sum of randomly focused Gaussian waves~\cite{Colaitis14,Colaitis15a,Ruocco19}, these results may be immediately applicable.

\section*{Acknowledgments}
The authors thank P.~Michel and T.~Chapman for helpful conversations. This work was performed under the auspices of the U.S. Department of Energy by Lawrence Livermore National Laboratory under Contract DE-AC52-07NA27344.

\section*{Disclosures}

The authors declare no conflicts of interest.

\appendix


\section{Hyperbolic umbilic catastrophe function}

\label{app:Humbilic}

The standard $D_4^+$ hyperbolic umbilic catastrophe function is defined in \Ref{Poston96} as
\begin{align}
	&\hUMB\left(
		t_1, t_2, t_3
	\right)
	\doteq
	\int \dd u \, \dd v
	\,
	\exp \left(
		i u^2 v
		+ i v^3
		+ i t_3 u^2
		\right.\nonumber\\
		&\hspace{52mm}\left.
		+ i t_2 u
		+ i t_1 v
	\right)
	.
	\label{eq:postonHU}
\end{align}

\noindent By making the variable substitution 
\begin{equation}
	u = 
	\sqrt{3} \,
	\frac{ \nu - \mu }{ 2^{2/3}}
	, \quad
	v = 
	\frac{\nu + \mu}{2^{2/3}} - \frac{t_3}{2}
	,
\end{equation}

\noindent one can also transform $\hUMB$ into the form
\begin{align}
	&\hUMB\left( t_1, t_2, t_3 \right)
	=
	\frac{\sqrt{3} }{ \sqrt[3]{2} }
	\exp\left(
		- \frac{i}{2} \, t_1 t_3 - \frac{i}{8} \, t_3^3
	\right)
	\nonumber\\
	&\fourier{U}_H \left(
		\frac{3 \, t_3^2 + 4 \, t_1 - 4 \sqrt{3} \, t_2}{2^{8/3}}
		,
		\frac{3 \, t_3^2 + 4 \, t_1 + 4 \sqrt{3} \, t_2}{2^{8/3}}
		,
		- \frac{3 \, t_3}{\sqrt[3]{2}}
	\right)
	,
\end{align}

\noindent where $\fourier{U}_H$ is the symmetrized hyperbolic umbilic catastrophe function used by \Ref{Olver10a}:
\begin{align}
	&\fourier{U}_H\left( \tau_1, \tau_2, \tau_3 \right)
	\doteq
	\int \dd \nu \, \dd \mu \,
	\exp \left(
		i \nu^3
		+ i \mu^3
		+ i \tau_3 \, \nu \mu
		\right.\nonumber\\
		&\hspace{50mm}\left.
		+ i \tau_2 \, \nu
		+ i \tau_1 \, \mu
	\right)
	.
\end{align}

\noindent Additionally, the integration over $v$ in \Eq{eq:postonHU} can be explicitly performed to yield the representation
\begin{align}
	\hUMB(t_1, t_2, t_3)
	&= \frac{2 \pi}{ \sqrt[3]{3} }
	\int
	\dd u \,
	\airyA\left(
		\frac{u^2 + t_1}{ \sqrt[3]{3} }
	\right)
    \nonumber\\
    &\hspace{20mm}\times
	\exp \left(
		i t_3 u^2
		+ i t_2 u
	\right)
	.
    \label{eq:airyHUMB}
\end{align}

\noindent Similarly, one can perform the Gaussian integration over $u$ in \Eq{eq:postonHU} and then make the variable transformation $z = \sqrt{v + t_3}$ to obtain the representation
\begin{align}
    &\hUMB(t_1, t_2, t_3)
    =
    2 \sqrt{i \pi}
    \exp\left(
        - i t_3^3
        - i t_1 t_3
    \right)
    \nonumber\\
    &\times
    \int_{i \infty}^\infty \dd z \,
    \exp\left[
        i z^6
        - 3 i t_3 z^4
        + i (3 t_3^2 + t_1) z^2 
        - i \frac{t_2^2}{4z^2}
    \right]
    ,
    \label{eq:hUMBvint}
\end{align}

\noindent where the integration contour passes from $+i \infty$ on the imaginary axis towards the origin, passing to the upper right of the essential singularity at $z = 0$, then continuing towards $+\infty$ on the real line. Equation \eq{eq:airyHUMB} is useful for numerical computation, as discussed further in \App{app:compute}, while \Eq{eq:hUMBvint} is useful for understanding the (nontrivial) asymptotic behavior of $\hUMB$ via steepest-descent methods, as discussed further in \Ref{Berry90b}.

Important for our purposes are the caustic surfaces of $\hUMB$. At fixed $t_3 \neq 0$, these caustics consist of one fold line and one cusp line in the $t_1-t_2$ plane, given by the parametric equations
\begin{subequations}
    \label{eq:causticTFOLD}
	\begin{align}
		t_1^{\text{(fold)}} &=
		- \frac{3}{2} t_3^2 \cosh(s)
		\left[
			\cosh(s) - 1
		\right]
		, \\
		t_2^{\text{(fold)}} &=
		\frac{\sqrt{3}}{2} t_3^2 \sinh(s)
		\left[
			\cosh(s) + 1
		\right]
		,
	\end{align}
\end{subequations}

\noindent and
\begin{subequations}
	\begin{align}
	    t_1^{\text{(cusp)}} &=
		- \frac{3}{2} t_3^2 \cosh(s)
		\left[
			\cosh(s) + 1
		\right]
		, \\
		t_2^{\text{(cusp)}} &=
		\frac{\sqrt{3}}{2} t_3^2 \sinh(s)
		\left[
			\cosh(s) - 1
		\right]
		,
	\end{align}
\end{subequations}

\noindent where the parameterization $s$ ranges from $-\infty$ to $\infty$. When $t_3 = 0$, the fold and cusp lines coalesce into the line segments
\begin{subequations}
    \label{eq:causticTSEG}
	\begin{align}
		t_1 \le 0
		, \quad
		t_2 &= \frac{t_1}{ \sqrt{3} }
		, \\
		t_1 \le 0
		, \quad
		t_2 &= - \frac{t_1}{ \sqrt{3} }
		.
	\end{align}
\end{subequations}

\noindent For $t_2 \approx 0$ (equiv., $s \approx 0$) and $t_3 \neq 0$, the fold and cusp lines are approximately represented as
\begin{subequations}
    \begin{align}
        \label{eq:causticTFOLDapprox}
    	t_1^\text{(fold)} &\approx
	    - \left( \frac{t_2}{2 t_3} \right)^2
    	, \\
	    t_1^\text{(cusp)} &\approx
    	- 3 t_3^2
	    - 9 \left( \frac{t_2 t_3}{2 \sqrt{3}} \right)^{2/3}
    	.
    \end{align}
\end{subequations}

\noindent Hence, the characteristic fold-line width, cusp-line width, and fold-cusp separation are respectively $t_3$, $1/t_3$, and $t_3^2$.

Since the fold and cusp lines become increasingly separated as $|t_3|$ increases, one expects that in the asymptotic limit $|t_3| \to \infty$ the hyperbolic umbilic function can be approximately represented as an Airy function. Indeed, by completing the square in \Eq{eq:airyHUMB}, one can represent $\hUMB$ as
\begin{align}
	\hUMB(t_1, t_2, t_3) &= \frac{2\pi}{\sqrt[3]{3}}
	\exp\left(
		- i \frac{t_2^2 }{4 t_3}
	\right)
	\int \dd u \,
	\airyA\left(
		\frac{u^2 + t_1}{\sqrt[3]{3}}
	\right)
    \nonumber\\
    &\hspace{4mm}\times
	\exp\left[
		i t_3 \left(
			u + \frac{t_2}{2 t_3}
		\right)^2
	\right]
	.
\end{align}

\noindent As $|t_3| \to \infty$, the integral will thus be dominated by the contributions around the stationary point $u = - t_2/(2 t_3)$. Hence, standard stationary phase methods yield the approximation
\begin{align}
	\hUMB(t_1, t_2, t_3) &\approx \frac{2\pi}{\sqrt[3]{3}}
	\sqrt{\frac{i \pi}{t_3} }
	\airyA\left[
		\frac{
			t_1
            + (t_2/2 t_3)^2
		}{\sqrt[3]{3}}
	\right]
    \nonumber\\
    &\hspace{4mm}\times
	\exp\left(
		- i \frac{ t_2^2 }{4 t_3}
	\right)
	.
	\label{eq:UHairy}
\end{align}

\noindent Note that the argument to the Airy function is simply the quadratic approximation to the fold line presented in \Eq{eq:causticTFOLDapprox}; hence, the basic structure of \Eq{eq:UHairy} can be anticipated from the principles of catastrophe optics.


\section{Ray equations for visualizing the caustic skeleton of $\hUMB$}

\label{app:rays}

In the following, we neglect the dissipation and finite beamwaist terms such that all physical parameters are real. For the wave equation given in \Eq{eq:airyEQ}, the dispersion symbol~\cite{Tracy14} that governs the propagation of the geometrical-optics rays is calculated to be
\begin{equation}
	\Symb{D}(x, y, k_x, k_y)
	= 
	k_x^2 + k_y^2
	+ \frac{x - \airyLEN}{\airySKIN^3}
	.
\end{equation}

\noindent The rays then satisfy the dynamical equations
\begin{subequations}
    \label{eq:rayEQs}
    \begin{align}
	    \pd{\tau} x(\tau) &= 2 k_x(\tau) 
    	, \quad
	    \pd{\tau} k_x(\tau) = - \frac{1}{\airySKIN^3}
    	, \\
	    \pd{\tau} y(\tau) &= 2 k_y(\tau)  
    	, \quad
	    \pd{\tau} k_x(\tau) = 0
    	.
    \end{align}
\end{subequations}

\noindent Let us normalize all wavenumber-like quantities by the vacuum wavenumber $2\pi/\lambda$, all distance-like quantities by the density lengthscale $\airyLEN$, and the ray propagation time (which has units of length squared) by their product:
\begin{subequations}
    \label{eq:normCOORD}
    \begin{align}
    	\tilde{x} &= \frac{x}{\airyLEN}
	    , \quad
    	\hspace{5mm}\tilde{y} = \frac{y}{\airyLEN}
	    , \quad
    	\hspace{3mm}\tilde{\focal} = \frac{\focal}{\airyLEN}
	    , \\
    	\tilde{k}_x &= \frac{\lambda k_x}{2\pi}
	    , \quad
    	\tilde{k}_y = \frac{\lambda k_y}{2\pi}
	    , \quad
    	\tilde{\tau} = \frac{2\pi \tau}{\lambda \airyLEN}
	    .
    \end{align}
\end{subequations}

\noindent Let us choose $\tilde{x}(0) = 0$ and $\tilde{y}(0) = \tilde{y}_0$. The initial condition \eq{eq:psiINhumb} for $\hUMB$ implies that the rays have the corresponding initial condition
\begin{equation}
	\tilde{k}_y(0)
	= \sin \theta - \frac{\tilde{y_0}}{\tilde{\focal}} \cos^2 \theta
	.
    \label{eq:ky0}
\end{equation}

\noindent The condition $\Symb{D} = 0$ then determines the remaining initial condition:
\begin{equation}
	\tilde{k}_x(0) = \sqrt{ 1 - \tilde{k}_y^2(0)}
	.
\end{equation}

\noindent The normalized ray trajectories that satisfy \Eq{eq:rayEQs} subject to the initial conditions are then given as
\begin{subequations}
    \label{eq:rays}
    \begin{align}
	    \tilde{x}(\tilde{\tau}) &= 2 \tilde{k}_x(0) \tilde{\tau} - \tilde{\tau}^2
    	, \quad
	    \tilde{k}_x(\tilde{\tau}) = \tilde{k}_x(0) - \tilde{\tau}
    	, \\
	    \tilde{y}(\tilde{\tau}) &= \tilde{y}_0 + 2 \tilde{k}_y(0) \tilde{\tau}
    	, \quad
	    \tilde{k}_y(\tilde{\tau}) = \tilde{k}_y(0)
    	.
    \end{align}
\end{subequations}

Note that the ray equations \eq{eq:rays} only describe the hyperbolic umbilic caustic pattern close to the critical point due to us applying an asymptotic initial condition at a finite location. Aberrations manifest farther from the critical point that cause the ray envelope caustic to deviate from the true caustic, although some authors choose to accommodate such aberrations in their unfolding convention for $\hUMB$ (equivalently, generically consider observations on curved surfaces rather than planes) to simplify the process of identifying this caustic in real experiments (see, for example, Figs.~4.4 and 4.5 in \Ref{Nye99}).


\section{Numerical procedure for computing $\hUMB$ and related functions}
\label{app:compute}

Here we provide a simple procedure for computing $\hUMB(t_1, t_2, t_3)$ and related integrals using the representation provided by \Eq{eq:airyHUMB}. First, when $t_3 = 0$ one can use results from \Ref{Vallee97} to obtain the exact expression 
\begin{align}
	\hUMB(t_1, t_2, 0)
	&=
    2 \pi^2
    \sqrt[6]{
        \frac{ 16}{3}
    }
	\airyA\left(
		\frac{
			t_1 
			- \sqrt{3} \, t_2
		}{
			\sqrt[3]{12}
		}
	\right)
    \nonumber\\
    &\hspace{15mm}\times
	\airyA\left(
		\frac{
			t_1 
			+ \sqrt{3} \, t_2
		}{
			\sqrt[3]{12}
		}
	\right)
	.
\end{align}

\noindent For general $t_3 \neq 0$, \Eq{eq:airyHUMB} takes the form of an FT:
\begin{equation}
	\hUMB(t_1, t_2, t_3)
	= \frac{2 \pi}{ \sqrt[3]{3} }
	\mc{F}_u
	\left[
		\airyA\left(
			\frac{u^2 + t_1}{ \sqrt[3]{3} }
		\right)
		\exp \left(
			i t_3 u^2
		\right)
	\right](t_2)
	,
	\label{eq:numHU}
\end{equation}

\noindent where $\mc{F}_u\left[ f(u) \right](k)$ denotes the FT of $f(u)$, now considered a function of $k$ defined as
\begin{equation}
	\mc{F}_u\left[ f(u) \right](k)
	\doteq
	\int
	\dd u \,
	f(u)
	\exp(-i k u)
	.
\end{equation}

\noindent Equation \eq{eq:numHU} can then be computed using standard FFT routines, which is how \Fig{fig:HUmbilicFIELD} was produced.

The apertured formulas of \Sec{sec:aper} can be computed similarly. Equation \eq{eq:planeAPER} is computed via FFT as
\begin{equation}
    \psi \propto
    \frac{\widetilde{\width}}{ \pi }
	\mc{F}_u
	\left[
		\sinc(\widetilde{\width} u)
        \airyA\left(
			\frac{u^2 + t_1}{ \sqrt[3]{3} }
		\right)
		\exp \left(
			i t_3 u^2
		\right)
	\right](t_2)
    ,
\end{equation}

\noindent where $t_1$, $t_2$, and $t_3$ are the arguments for $\hUMBKER$ in \Eq{eq:planeAPER}, while \Eq{eq:focusAPER} is computed as
\begin{equation}
    \psi \propto
    \frac{\pi}{\sqrt[3]{3}}
	\mc{F}_u
	\left[
		\aperFUNC(u)
        \airyA\left(
			\frac{u^2 + t_1}{ \sqrt[3]{3} }
		\right)
		\exp \left(
			i t_3 u^2
		\right)
	\right](t_2)
    ,
\end{equation}

\noindent where $t_1$, $t_2$, and $t_3$ are the arguments for $\hUMBKER$ in \Eq{eq:focusAPER}. Figures \ref{fig:planeAPER} and \ref{fig:focusAPER} are computed using these formulas.

\bibliography{Biblio.bib}
\bibliographystyle{apsrev4-1}

\end{document}